\documentclass[lettersize,journal]{IEEEtran}
\usepackage{cite}
\usepackage{amsmath,amssymb,amsfonts}
\usepackage{graphicx,color}
\usepackage{textcomp}

\usepackage{amsmath,amsfonts}
\usepackage{array}

\usepackage{textcomp}
\usepackage{stfloats}
\usepackage{url}
\usepackage{verbatim}
\usepackage{graphicx}
\usepackage{cite}
\usepackage{xcolor}
\usepackage{hyperref}
\usepackage{multirow}

\usepackage{caption}
\usepackage{subcaption}
\usepackage{romannum}
\usepackage{fix-cm}

\usepackage{booktabs} 

\usepackage{pifont} 

\newcommand{\cmark}{\ding{51}} 
\usepackage{makecell} 

\usepackage[linesnumbered,ruled,vlined]{algorithm2e}

\usepackage{algpseudocode}

\begin{document}




\title{Multi-Task Self-Supervised Learning for Generalizable IQ Representations: Toward Foundation Models for AI-Native 6G}

\title{Toward IQ-based 6G AI-Native Foundation Models using Multi-Task Self-Supervised Learning}

\title{Toward 6G AI-Native Foundation Models using Self-Supervised Learning on  IQ-Representations}

\title{Toward AI-Native 6G Foundation Models using Self-Supervised Representation Learning of \\MIMO IQ-Streams}

\title{From Raw IQ Data to Foundation Models: A Step Toward AI-Native 6G}

\title{Enabling Foundation Models for AI-Native 6G from Raw MIMO IQ Data}
\title{IQFM: A Foundational Model for Wireless Tasks Using Raw IQ Data}

\title{\textbf{6G \textsc{native}FM:} A Wireless Foundational Model for I/Q Signals} 

\title{\textbf{I/Q \textsc{native}FM:} A 6G Wireless Foundational Model for I/Q Signals} 

\title{\textbf{ \textsc{i/q}FM:} A  NativeAI 6G Foundational Model for I/Q Signals}

\title{\textbf{ \textsc{i/q}FM:} A Wireless Foundational Model for AI-Native 6G Networks}

\title{\textbf{ \textsc{iq}FM}- 
A Wireless Foundational Model for I/Q Streams in AI-Native 6G}

\title{\textbf{{\fontsize{20}{9.5}\selectfont IQ}FM} -- A Wireless Foundational Model \\for I/Q Streams in AI-Native 6G}





\author{
    \IEEEauthorblockN{
        Omar Mashaal and Hatem Abou-Zeid \\
        WAVES LAB, Department of Electrical and Software Engineering, University of Calgary, Calgary, Canada \\
        \ttfamily \{omar.mashaal1,hatem.abouzeid\}@ucalgary.ca
    }
}



\maketitle
\begin{abstract}
Foundational models have shown remarkable potential in natural language processing and computer vision, yet remain in their infancy in wireless communications. While a few efforts have explored image-based modalities such as channel state information (CSI) and frequency spectrograms, foundational models that operate directly on raw IQ data remain largely unexplored. This paper presents, IQFM, the first I/Q signal foundational model for wireless communications. IQFM supporting diverse tasks: modulation classification, angle-of-arrival (AoA), beam prediction, and RF fingerprinting, without heavy preprocessing or handcrafted features. We also introduce a task-aware augmentation strategy that categorizes transformations into core augmentations, such as cyclic time shifting, and task-specific augmentations. This strategy forms the basis for structured, task-dependent representation learning within a contrastive self-supervised learning (SSL) framework. Using this strategy, the lightweight encoder, pre-trained via SSL on over-the-air multi-antenna IQ data, achieves up to 99.67\% and 65.45\% accuracy on modulation and AoA classification, respectively, using only one labeled sample per class, outperforming supervised baselines by up to 7$\times$ and 145$\times$. The model also generalizes to out-of-distribution tasks; when adapted to new tasks using only 500 samples per class and minimal parameter updates via LoRA, the same frozen encoder achieves 94.15\% on beam prediction (vs. 89.53\% supervised), 50.00\% on RML2016a modulation classification (vs. 49.30\%), and 96.05\% on RF fingerprinting (vs. 96.64\%). These results demonstrate the potential of raw IQ-based foundational models as efficient, reusable encoders for multi-task learning in AI-native 6G systems.
\end{abstract}


\maketitle

\section{INTRODUCTION}

\IEEEPARstart{W}{ireless} communication systems are increasingly adopting machine learning (ML) methods to improve key tasks such as modulation classification\cite{mod_snr}, channel estimation \cite{chan_Est} and Angle-of-Arrival (AoA) prediction \cite{aoa_1,mallioras2024enhancing}. These data-driven methods improve signal classification under low signal-to-noise ratio (SNR) conditions and enhance localization accuracy, thereby facilitating more efficient spectrum utilization and interference management in applications like cognitive radio and adaptive beamforming \cite{ML_radio}. As wireless networks evolve toward sixth-generation (6G) architectures, which are envisioned as AI-native systems \cite{saad2024artificial,brik2022deep}, deep learning (DL) models are expected to play a central role in enabling intelligent spectrum allocation, real-time adaptation to channel dynamics, and autonomous decision-making. Despite the promising performance of supervised DL approaches, their dependency on large-scale labeled datasets presents limitations in terms of scalability, domain generalization, and practical deployment. In particular, the acquisition of labeled data is often costly and infeasible in dynamic wireless environments \cite{device_id}. This  emphasizes the need for more efficient learning paradigms that reduce reliance on manual annotations

Self-supervised learning (SSL) has emerged as a promising alternative for wireless communication tasks by enabling models to learn from unlabeled data. SSL eliminates the need for large annotated datasets and has shown success in specific wireless applications such as modulation classification~\cite{r_mod_1,r_mod_2,r_mod_3}, beam management~\cite{r_other_1}, and emitter identification~\cite{r_id}. However, existing work is typically limited to single-task scenarios, with models tailored to specific downstream objectives. As a result, current SSL approaches in raw IQ data modality do not fully exploit the potential for unified, general-purpose representation learning, a property that makes SSL a natural foundation for building foundational models.

Foundational models (FMs) refer to large, pre-trained models that learn generalizable representations from diverse, unlabeled data and can be adapted to a wide range of downstream tasks. They have demonstrated remarkable success in fields like computer vision~\cite{vit} and natural language processing~\cite{text}, and are beginning to gain interest in wireless communications. However, most existing wireless FMs rely on preprocessed features such as Channel State Information (CSI) or spectrograms~\cite{r_other_6,r_other_7,r_other_8}, which may discard important low-level signal information and add processing overhead. In contrast, this work presents, to the best of our knowledge, the first foundational model trained directly on raw IQ data. By avoiding handcrafted features, our approach enables more expressive, task-agnostic representations and supports multi-task learning across diverse wireless tasks such as RF fingerprinting, beam prediction, modulation classification, and angle-of-arrival prediction, facilitating low, latency inference in AI-native 6G systems.

To address this, we introduce an SSL framework designed specifically for unprocessed multi-channel MIMO IQ data. This framework supports efficient multi-task learning across diverse wireless tasks by preserving the unique signal characteristics necessary for each task. The key contributions of this work are summarized as follows:
\begin{figure*}[t]
    \centering
    \includegraphics[trim=3cm 2.5cm 3.5cm 0.7cm, clip,width=0.9\linewidth]{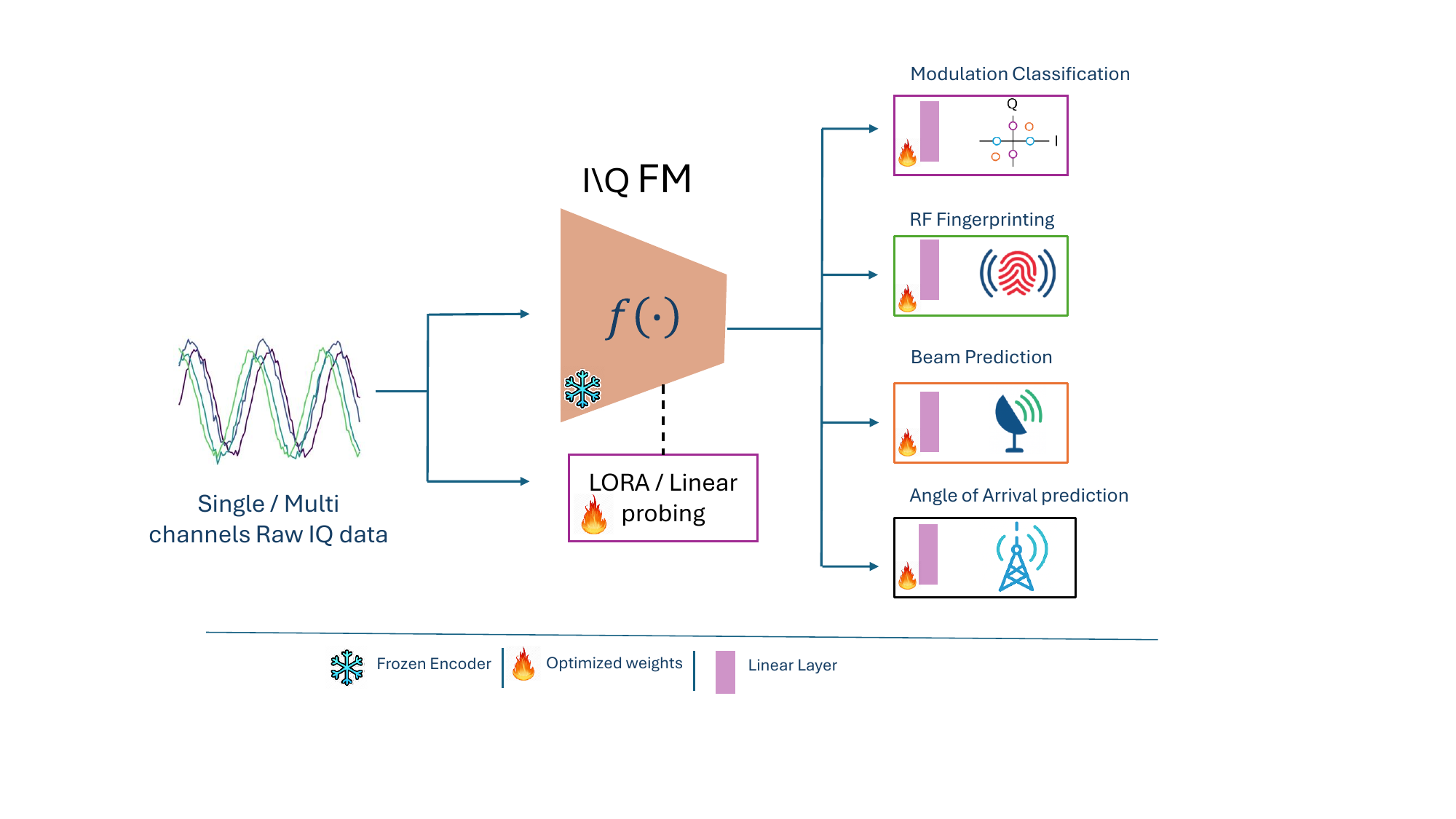}
    \caption{Proposed foundational model for raw IQ data: a shared encoder trained with SSL, adapted to multiple wireless tasks using lightweight fine-tuning strategies.}
    \label{fig:framework}
\end{figure*}

\begin{itemize}

\item We present, to the best of our knowledge, the first foundational model for wireless communication signals that operates directly on raw IQ data.
The model supports a variety of tasks, including modulation and angle-of-arrival classification, RF fingerprinting, and beam prediction, demonstrating multi-task learning capabilities across diverse wireless tasks. Our architecture employs a lightweight ShuffleNetV2 (0.5×) backbone, enabling efficient feature extraction and real-time inference while maintaining a compact model footprint suitable for resource-constrained deployment.

\item We investigate lightweight fine-tuning strategies for adapting the foundational model to both in-distribution datasets and entirely new tasks. Our results show that linear probing is effective for multi-task adaptation on in-distribution data, achieving strong performance in the few-shot regime. For out-of-distribution tasks and unseen datasets, we demonstrate that parameter-efficient Low-Rank Adaptation (LoRA) enables robust and lightweight adaptation, achieving competitive results across a range of label budgets while introducing minimal additional parameters compared to full model fine-tuning.

\item We introduce a principled augmentation strategy for raw MIMO IQ data that separates core augmentations, such as cyclic time shifting, which preserve features across tasks, from task-specific augmentations designed to emphasize distinct signal dependencies (e.g., channel dropping for modulation and antenna masking for AoA). To our knowledge, this is the first application of cyclic time shifting in multi-task SSL for raw MIMO IQ data, and the first to organize augmentations based on signal-domain relevance and task-specific utility.

\item We validate the foundational model’s practical effectiveness through comprehensive evaluations on multiple over-the-air datasets. On the in-distribution testbed dataset, task-specific SSL improves modulation accuracy from 14.27\% to 99.67\% ( $\sim$7$\times$ ) and AoA accuracy from 0.45\% to 65.45\% ( $\sim$145$\times$ ) with only one labeled sample per class. The joint SSL model achieves 95.71\% and 89.35\% accuracy for modulation and AoA, respectively, with just 10 labeled samples per class, outperforming fully supervised baselines by over 2.3$\times$ and 1.3$\times$. The model also generalizes effectively to out-of-distribution tasks, achieving 66.9\% accuracy with a single labeled sample per class on RF fingerprinting (2.5$\times$ supervised), 52.6\% at 50 samples per class on beam prediction (1.2$\times$ supervised), and 38.1\% at 50 samples per class on RML2016 modulation classification (1.5$\times$ supervised).
\end{itemize}

Overall, this work advances the development foundational models for AI-native 6G wireless systems, enabling low-overhead, real-time inference across diverse wireless tasks directly from raw IQ data. The rest of the paper is organized as follows: Section~\Romannum{2} reviews related work. Section~\Romannum{3} presents the signal model and problem formulation. Section~\Romannum{4} discusses the experimental results and key findings. Finally, Section~\Romannum{5} concludes the paper.

\section{Related work}

\subsection{Single-Task Self-Supervised Learning in Wireless}

Several works have employed SSL for modulation classification by training on raw IQ data or transformed signal representations. In \cite{r_mod_1}, a ResNet-50 encoder was trained using contrastive learning and later fine-tuned. Transformer-based architectures were explored in \cite{r_mod_2}, while Self-RadioNet \cite{r_mod_3} and SemiAMC \cite{r_mod_4} proposed improvements to contrastive pipelines tailored for modulation tasks. A debiased hardness-weighted loss was introduced in \cite{r_mod_5} to better separate challenging samples. Beyond raw data, some approaches utilize image-based signal formats such as spectrograms \cite{r_mod_6} and Gramian Angular Field images \cite{r_mod_7}.

SSL has also been applied to spatial and channel-related tasks using preprocessed inputs like CSI and Channel Impulse Response (CIR). In \cite{r_other_1}, CSI-derived features were used for mmWave beamforming, while \cite{r_other_2} learned geolocation-aware embeddings from MIMO-OFDM data. In RIS-assisted MIMO, SSL was employed for joint beamforming optimization \cite{r_other_3}. For MIMO radar, \cite{r_other_4} enhanced angular resolution by predicting antenna responses from range-Doppler-antenna data.

While effective, these works remain task-specific and lack a unified framework that supports multiple downstream wireless applications from shared representations.

\subsection{Multi-Task Learning in Wireless Communication}

Multi-task learning (MTL) has been investigated in wireless communication to enable the joint modeling of related tasks. In \cite{r_other_5}, a fully supervised MTL model was proposed that operates on raw IQ data for simultaneous modulation and protocol classification, utilizing a shared encoder with task-specific output heads. However, their approach remains fully supervised and does not incorporate self-supervised learning principles. While self-supervised multi-task learning has been explored using preprocessed representations such as CSI or CIR \cite{r_other_6}, efforts to extend SSL directly to raw IQ signals are still limited.

The work in \cite{kanu2025ssl} represents the first attempt to apply multi-task SSL on raw MIMO IQ data for joint modulation classification and AoA estimation. Their results showed the feasibility of learning shared representations across tasks using contrastive learning. However, strong performance required encoder fine-tuning and thousands of labeled examples, indicating the need for more efficient frameworks under stricter low-label conditions

Building on this foundation, our work introduces two key advancements; First, we incorporate cyclic time shifting as a core augmentation to support feature learning across tasks, an aspect not explored in \cite{kanu2025ssl}. Second, we develop a task-aware augmentation framework that separates transformations common to multiple tasks from those specific to individual task characteristics. This structured link between augmentation design and task requirements was not considered in \cite{kanu2025ssl}. These contributions enable efficient low-shot multi-task learning from raw IQ data without requiring encoder retraining.

\subsection{Foundational Models for Wireless Communication}

Inspired by the success of large-scale pretraining in vision and language, foundation models are gaining traction in wireless communication, particularly in the context of emerging 6G and integrated sensing and communication (ISAC) systems. WavesFM \cite{r_other_7} builds upon earlier FM work presented by the same authors in \cite{other_20,r_other_19}, where they introduced and subsequently improved the first wireless FM leveraging SSL to extract generalizable representations from RF spectrograms. Their work demonstrated effective performance in multi-task scenarios, including human activity sensing and spectrogram segmentation, using a single shared FM. WavesFM extends these approaches further by adopting a ViT-based architecture with task-specific heads and LoRA, employing masked modeling techniques on spectrograms and CSI, trained on real-world data to support a broader spectrum of communication, sensing, and localization tasks.

Wireless channel representation learning has also gained attention in recent literature. CSI-CLIP \cite{r_other_8} introduced a contrastive learning method, treating CSI and CIR as aligned modalities to pretrain a wireless channel foundation model, facilitating tasks like positioning, beam management, and line-of-sight (LOS) versus non-line-of-sight (NLOS) identification. Similarly, WirelessGPT \cite{r_other_6} employed a masked reconstruction strategy to learn robust representations from extensive simulated datasets, enabling effective performance in channel prediction and human activity recognition tasks. Both methods primarily rely on simulated data during pre-training and fine-tuning phases.

While these models demonstrate versatility, they typically rely on structured inputs such as CSI, CIR, or spectrograms, which require extensive preprocessing. In contrast, foundational models that operate directly on raw IQ data remain largely unexplored. Our work addresses this gap by proposing a raw-IQ-based FM that enables multi-task learning without handcrafted features and generalizes to unseen downstream tasks, offering improved adaptability across domains



\section{Signal Model and Problem Formulation}

We consider wireless signals captured using multiple antennas, arranged in a uniform linear array (ULA) with \( M \) antenna elements, spaced uniformly by a distance \( d \). The position of the \( m \)-th antenna element is defined as:
\begin{equation}
p_m = (m-1)\,d, \quad m \in \{1,\dots,M\}.
\end{equation}

Assuming narrowband and far-field conditions, the complex baseband signal received by antenna element \( m \), from a wave arriving at an angle-of-arrival (AoA) \( \theta \), is modeled as:
\begin{equation}
x_m(t) = \alpha\, s(t)\, e^{j 2\pi \frac{p_m}{\lambda} \sin(\theta)} + n_m(t),
\label{eq:ULA_SignalModel}
\end{equation}
where \( \alpha \) denotes a complex channel gain, \( s(t) \) is the transmitted waveform, and \( n_m(t) \) is additive noise.

In practical deployments, each antenna element samples the received signal over \( T \) discrete time steps, capturing both the in-phase (I) and quadrature (Q) components. This yields a real-valued tensor of shape:
\begin{equation}
\mathbf{X} \in \mathbb{R}^{M \times 2 \times T},
\end{equation}
where the second dimension distinguishes the I/Q channels.

Unlike single-channel IQ data, multi-antenna (MIMO) IQ data provides richer information by capturing both spatial and temporal signal structures. The spatial diversity across antenna elements encodes directional characteristics such as AoA. In parallel, the temporal structure of the received waveform reflects modulation-specific characteristics and device-level variations, which are critical for tasks like modulation classification and RF fingerprinting.

Our primary objective is to leverage these spatial and temporal structures to train a foundational encoder \( f(\cdot;\theta) \) that learns generalizable representations from raw MIMO IQ data. The encoder is pretrained using contrastive SSL, without reliance on handcrafted features or explicit domain-specific preprocessing.

After pretraining, the encoder is frozen and used as a shared feature extractor. Lightweight, task-specific heads \( h_t(\cdot) \) are trained using minimal labeled data for various downstream tasks such as modulation classification, AoA classification, RF fingerprinting, and beam prediction. Each task head is optimized by minimizing a cross-entropy (CE) loss:
\begin{equation}
\mathcal{L}_t = \mathrm{CE}\left(h_t(f(\mathbf{X})),\, y_t\right).
\end{equation}

This formulation enables parameter-efficient adaptation via linear probing or Low-Rank Adaptation, allowing a single foundational encoder to support multiple wireless signal understanding tasks with high efficiency and generalization.


\begin{figure*}[t]
    \centering
    \includegraphics[trim=0cm 2.8cm 0cm 3cm, clip,width=0.98\linewidth]{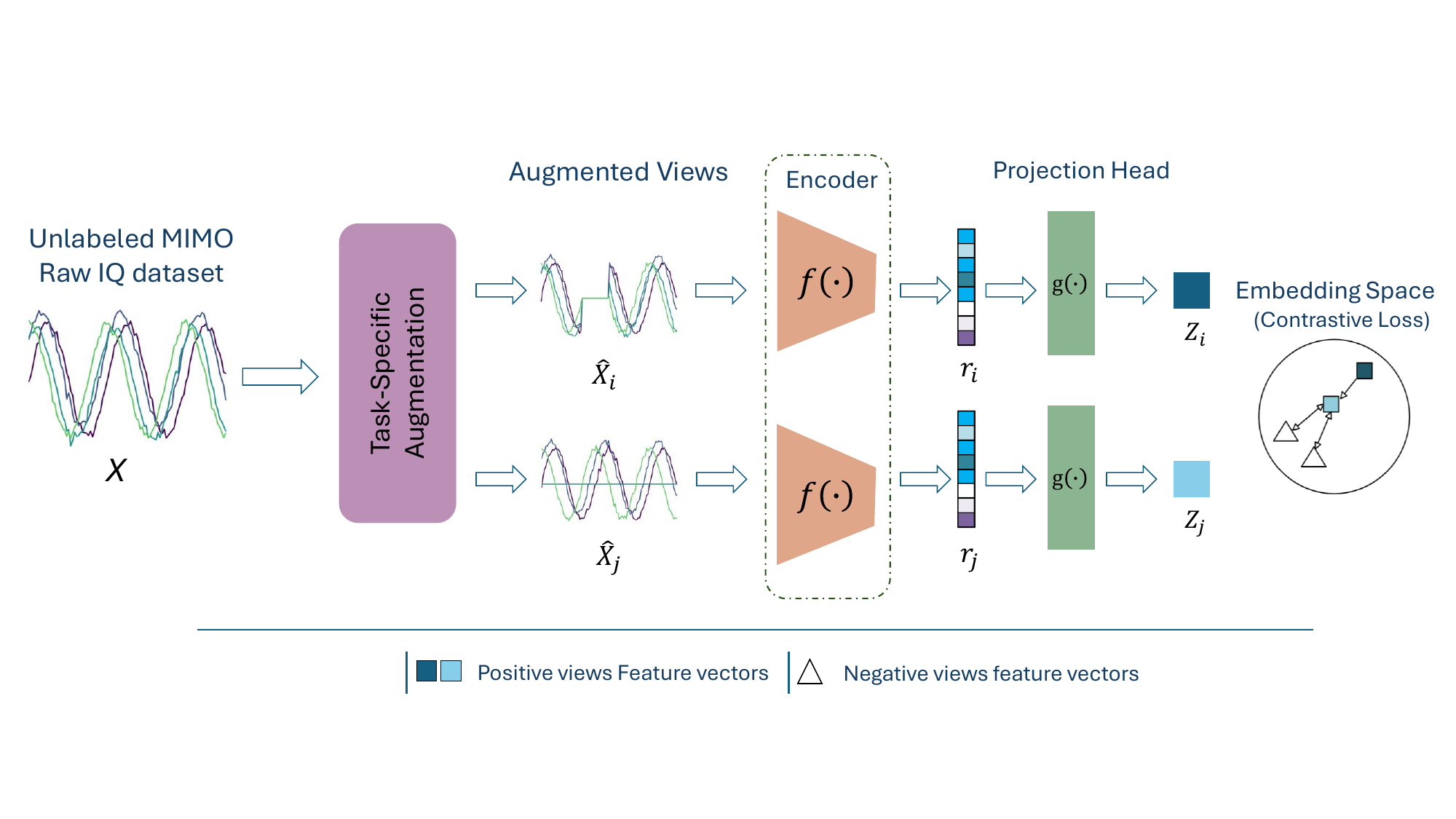}
    \caption{The proposed foundational model training for raw IQ data using contrastive learning and task-specific augmentations to capture temporal and spatial features.}
    \label{fig:frameworkmethod}
\end{figure*}

\section{Methodology}
\label{sec:methodology}
\subsection{ Overview of I/Q FM Framework}

The proposed foundational model, illustrated in Fig.~\ref{fig:framework}, consists of a shared encoder trained on unlabeled raw MIMO IQ samples. This encoder generates a general-purpose representation suitable for various wireless communication tasks. Adaptation for specific tasks is efficiently handled by lightweight linear probes or parameter-efficient fine-tuning via LoRA. During task-specific fine-tuning, the encoder is kept frozen to ensure low computational overhead and robust generalization capabilities.

\subsection{Contrastive Self-Supervised Pre-training}

To pre-train I/Q FM, we employ a contrastive SSL framework inspired by SimCLR~\cite{simclr}. The core objective of contrastive learning is to align representations of different augmented views of the same sample (positive pairs) while pushing apart representations of different samples (negative pairs). A key component of our framework is the use of task-aware augmentations tailored to the unique temporal and spatial structures present in raw MIMO IQ data. As illustrated in Figure~\ref{fig:frameworkmethod}, raw IQ samples undergo task-specific augmentations before encoding. These transformations are designed to promote the learning of representations that are both broadly transferable and specifically aligned with the corresponding signal properties of each task. While we adopt SimCLR as a baseline for contrastive SSL, the framework is compatible with other paradigms, such as MoCo~\cite{moco} and BYOL~\cite{byol}, and can be extended accordingly.

Let \( \mathbf{X} \in \mathbb{R}^{M \times 2 \times T} \) denote the input tensor of IQ samples collected from a ULA with \( M \) antennas over \( T \) time steps. Two augmentation functions \( t_i, t_j \in \mathcal{T} \) are randomly sampled and applied to generate two views of the same sample:
\begin{align}
    \tilde{\mathbf{X}}_i &= t_i(\mathbf{X}), 
    &
    \tilde{\mathbf{X}}_j &= t_j(\mathbf{X}).
\end{align}

These views are passed through a shared encoder \( f_\theta(\cdot) \), followed by a projection head \( g_\phi(\cdot) \), yielding embeddings:
\begin{align}
    \mathbf{Z}_i &= g_\phi(f_\theta(\tilde{\mathbf{X}}_i)), 
    &
    \mathbf{Z}_j &= g_\phi(f_\theta(\tilde{\mathbf{X}}_j)).
\end{align}

The model is trained using the InfoNCE loss:
\begin{equation}
\label{eq:infoNCE}
    \mathcal{L}_{\mathrm{SSL}}
    =
    -\sum_{\mathbf{X}}
    \log 
      \frac
        {\exp\left(\mathrm{sim}(\mathbf{Z}_i,\mathbf{Z}_j)/\tau\right)}
        {\sum_{\mathbf{Z}^-}
        \exp\left(\mathrm{sim}(\mathbf{Z}_i,\mathbf{Z}^-)/\tau\right)},
\end{equation}
where \( \mathrm{sim}(\cdot, \cdot) \) denotes cosine similarity and \( \tau \) is a temperature hyperparameter. All other samples in the batch are treated as negatives.

\subsection{Signal Augmentations for Feature Generalization}

Signal augmentation design affects representation learning quality in contrastive SSL, particularly in domains like MIMO wireless communications where signals contain diverse structures, including temporal and spatial phase dependencies. For instance, AoA prediction depends primarily on spatial phase relationships across antenna elements, whereas modulation classification is more sensitive to temporal and spectral variations. Consequently, task-agnostic augmentations may distort essential features and lead to suboptimal performance.

\begin{figure}[!t]
    \centering
    \begin{subfigure}[b]{0.48\linewidth}
        \centering
        \includegraphics[width=\linewidth]{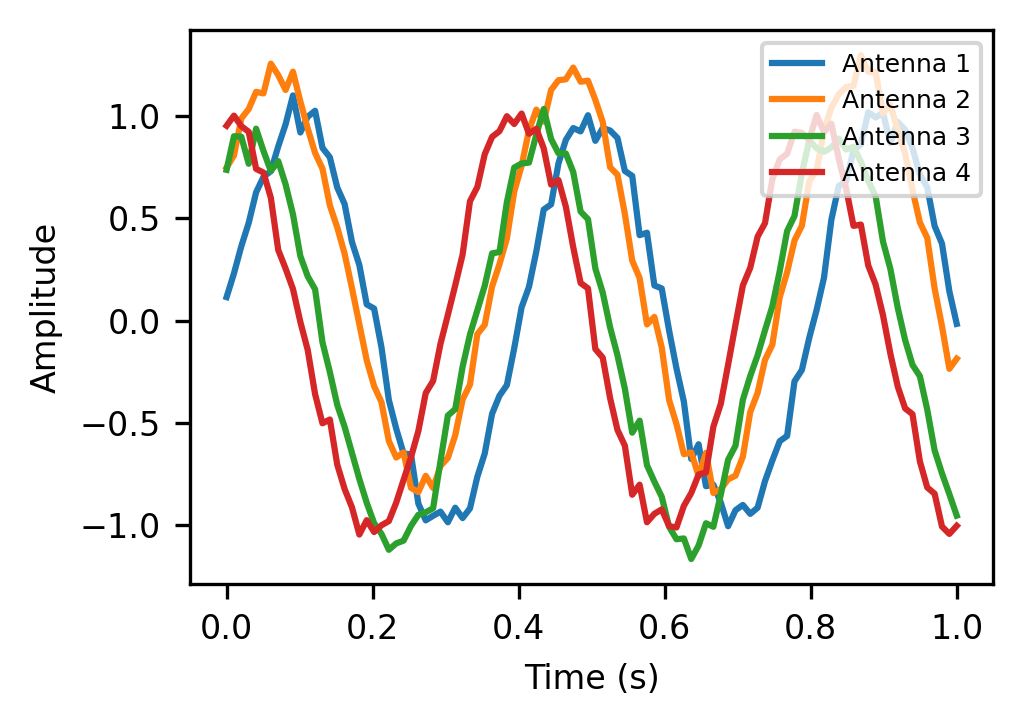}
        \caption{}
        \label{fig:original}
    \end{subfigure}
    \hspace{0.01\linewidth}
    \begin{subfigure}[b]{0.48\linewidth}
        \centering
        \includegraphics[width=\linewidth]{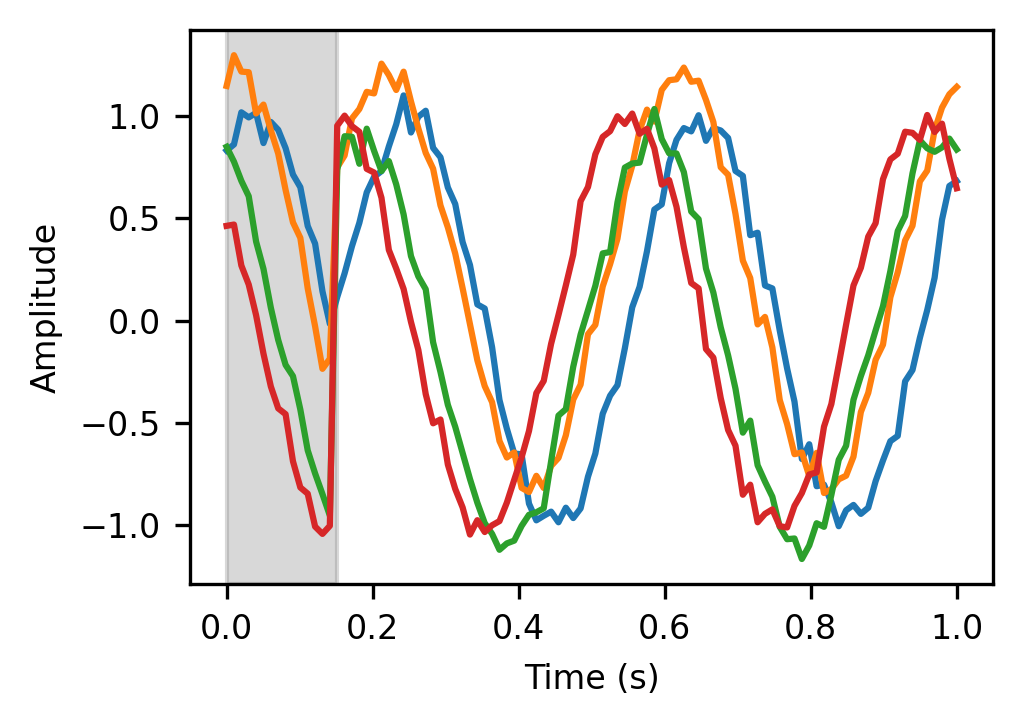}
        \caption{}
        \label{fig:time_rolling}
    \end{subfigure}

    \vspace{0.5em}

    \begin{subfigure}[b]{0.48\linewidth}
        \centering
        \includegraphics[width=\linewidth]{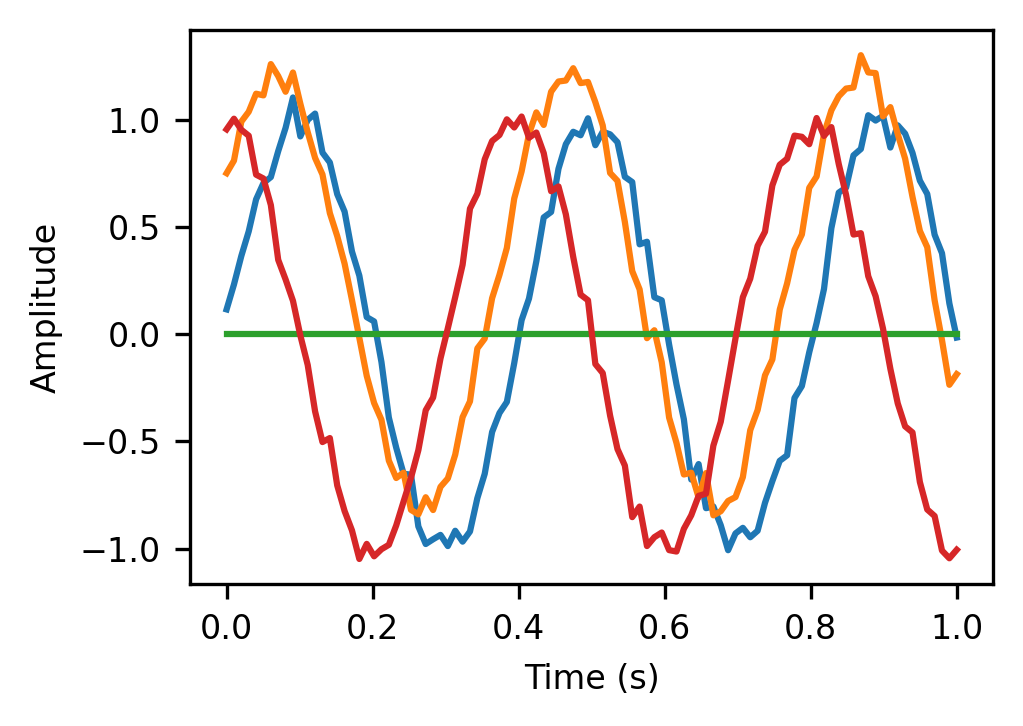}
        \caption{}
        \label{fig:channel_dropping}
    \end{subfigure}
    \hspace{0.01\linewidth}
    \begin{subfigure}[b]{0.48\linewidth}
        \centering
        \includegraphics[width=\linewidth]{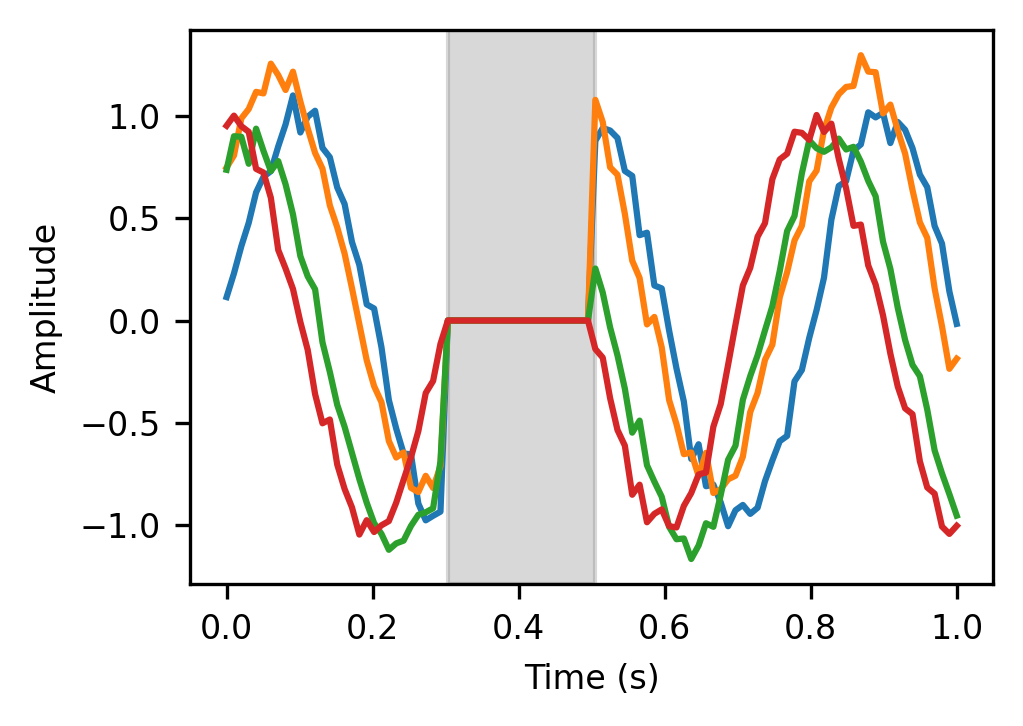}
        \caption{}
        \label{fig:channel_masking}
    \end{subfigure}

    \caption{Visualization of task-specific augmentations applied to a 4-antenna IQ signal: (a) original signal, (b) time rolling, (c) channel dropping, and (d) channel masking.}
    \label{fig:ssl_augmentations}
\end{figure}

We focus on three augmentations tailored for raw MIMO IQ signals. Cyclic time shift, introduced in this work and referred to as \emph{time rolling (TR)}, enhances temporal robustness while preserving inter-antenna phase structure, making it applicable to both tasks. Channel masking (CM) and channel dropping (CD), previously studied in related contexts, are revisited here with a task-aware perspective: CM preserves spatial structure and benefits spatially dependent tasks such as AoA, while CD introduces controlled temporal variation suited for tasks such as modulation classification. Figure~\ref{fig:ssl_augmentations} illustrates these transformations. In the subsequent section, we provide their formal definitions and analyze their respective impact on task-specific performance.

\subsection{Mathematical Analysis of Task-Specific Augmentations}
We illustrate the connection between each augmentation and the features it affects on MIMO IQ data by analyzing their impact on two tasks: AoA, which relies on spatial phase relationships across antennas, and modulation classification, which depends on temporal signal patterns.

\subsubsection{Cyclic Time-Shift Augmentation}

To simulate temporal misalignment, we apply a cyclic time shift (TR) along the time dimension, :

\begin{equation}
    \mathbf{X}_{\text{rolled}}(t) = \mathbf{X}((t + d \cdot \tau) \bmod T),
\end{equation}

where \( \tau \in \{0, \dots, T-1\} \) is a randomly selected shift amount, and \( d \in \{-1, 1\} \) is a randomly sampled direction, enabling both forward and backward rolling.

\paragraph{Effect on AoA Characteristics}

From the signal model, the received signal at the \( m \)-th antenna is given by:

\begin{equation}
    x_m(t) = \alpha s(t) e^{j \frac{2\pi}{\lambda} p_m \sin(\theta)} + n_m(t),
\end{equation}

Applying the cyclic time shift yields:

\begin{equation}
    x_m^{\text{shifted}}(t) = \alpha s(t + d \cdot \tau) e^{j \frac{2\pi}{\lambda} p_m \sin(\theta)} + n_m(t + d \cdot \tau).
\end{equation}

Assuming stationary noise, the spatial phase term \( e^{j \frac{2\pi}{\lambda} p_m \sin(\theta)} \) remains unchanged. Thus, the relative phase difference across antennas is preserved:

\begin{equation}
    \frac{x_m^{\text{shifted}}(t)}{x_n^{\text{shifted}}(t)} = e^{j \frac{2\pi}{\lambda} (p_m - p_n) \sin(\theta)}.
\end{equation}

This confirms that cyclic time shifting does not affect the spatial structure required for AoA prediction. As a result, the model is encouraged to learn spatially invariant features, specifically, inter-antenna phase differences, enhancing robustness to temporal shifts.

\paragraph{Impact on Modulation Characteristics}

Each modulation scheme exhibits characteristic temporal patterns and constellation dynamics. The cyclic time-shifting operation prevents the DL model from relying on specific time indices and encourages learning time-invariant features intrinsic to modulation.

According to the Fourier shift theorem, a time-domain shift results in a linear phase shift in the frequency domain:

\begin{equation}
    X_m^{\text{shifted}}(f) = e^{j 2\pi f (d \cdot \tau)} X_m(f),
\end{equation}

implying the magnitude spectrum remains unchanged:

\begin{equation}
    \left| X_m^{\text{shifted}}(f) \right| = \left| X_m(f) \right|.
\end{equation}

This confirms that cyclic time shifting preserves spectral features critical to modulation classification. Thus, the model learns discriminative features from modulation-specific properties, such as symbol transitions and constellation trajectories, rather than absolute time alignment.

Given that the self-supervised model operates directly on raw IQ samples, time-domain invariance is particularly important. The received signal at the \( m \)-th antenna is given by:
\begin{equation}
    x_m(t) = I_m(t) + j Q_m(t),
\end{equation}

where \( I_m(t) \) and \( Q_m(t) \) denote the in-phase and quadrature components, respectively. A cyclic time shift yields:

\begin{equation}
    x_m^{\text{shifted}}(t) = I_m((t + d \cdot \tau) \bmod T) + j Q_m((t + d \cdot \tau) \bmod T).
\end{equation}

This transformation preserves the signal's symbol structure and relative phase dynamics, leaving the statistical properties of modulation intact. Consequently, the augmentation improves model generalization by enforcing reliance on intrinsic modulation features rather than fixed temporal positions.

\subsubsection{Channel Masking Augmentation}

In this augmentation CM, a masking tensor $M \in \{0,1\}^{M \times 2 \times T}$ is applied to zero out specific time samples simultaneously across all antennas, preserving spatial coherence but introducing temporal sparsity:

\begin{equation}
    \mathbf{X}_{\text{masked}} = \mathbf{X} \odot M,
\end{equation}

where $\odot$ denotes element-wise multiplication.

\paragraph{Effect on AoA Characteristics}

Applying the channel masking on the received signal at antenna $m$, given by~\eqref{eq:ULA_SignalModel}, yields:

\begin{equation}
    x_m^{\text{masked}}(t) = M(t)\, x_m(t),
\end{equation}

where $M(t)$ is applied equally across all antennas. Thus, at unmasked time indices ($M(t)=1$), the relative phase difference across antennas remains intact:

\begin{equation}
    \frac{x_m^{\text{masked}}(t)}{x_n^{\text{masked}}(t)} 
    = \frac{x_m(t)}{x_n(t)}
    = e^{j \frac{2\pi}{\lambda}(p_m - p_n)\sin(\theta)}, \quad \text{for } M(t)=1.
\end{equation}

At masked indices ($M(t)=0$), no signal is observed. Thus, channel masking preserves the spatial relationships (phase differences) whenever signal information is available. Learning from this incomplete temporal data encourages the model to rely on spatially consistent features, enhancing robustness to missing temporal segments.

\paragraph{Impact on Modulation Characteristics}

Modulation classification relies on temporal symbol transitions and constellation patterns. Channel masking randomly zeros portions of the IQ signal, causing disruptions in these critical temporal structures:

\begin{equation}
    x_m^{\text{masked}}(t) = M(t)\, [I_m(t) + j Q_m(t)].
\end{equation}

Modulation classification relies heavily on temporal and spectral signal features. While CM disrupts these temporal structures by randomly removing signal portions, causing the model to rely more on spatial (antenna-phase) information. Consequently, modulation performance typically degrades, as spatial characteristics alone are insufficient for accurate modulation discrimination.


\subsubsection{Channel Dropping Augmentation}

In this augmentation CD, signals from a randomly selected subset of antennas are dropped (set to zero). Let $ D \subset \{1, \dots, M\} $ represent the set of dropped antennas. The augmented signal becomes:

\begin{equation}
    (\mathbf{X}_{\text{dropped}})_m = 
    \begin{cases} 
        \mathbf{X}_m, & \text{if } m \notin D, \\
        0, & \text{if } m \in D.
    \end{cases}
\end{equation}

\paragraph{Impact on Modulation Characteristics}

Modulation classification primarily relies on temporal features, including symbol transitions and frequency-domain patterns, within each antenna channel independently.

Applying channel dropping yields on the received given by~\eqref{eq:ULA_SignalModel}:

\begin{equation}
    x_m^{\text{dropped}}(t) = 
    \begin{cases} 
        x_m(t), & m \notin D, \\
        0, & m \in D.
    \end{cases}
\end{equation}

Since modulation information (symbol timing, transitions, spectral content) remains fully preserved within each active antenna independently, dropping antennas does not affect these intrinsic modulation features. As each antenna channel is processed separately, the model continues to learn robust modulation representations from the active channels, unaffected by spatial redundancy.

\paragraph{Effect on AoA Characteristics}

AoA estimation fundamentally relies on phase differences between multiple antennas:

\begin{equation}
    \Delta \phi_{m,n} = \frac{2\pi}{\lambda}(p_m - p_n)\sin(\theta).
\end{equation}

Dropping antennas removes their received signals, rendering phase information unavailable from those antennas. For example, if antenna \( j \) is dropped, its phase difference with an active antenna \( i \) becomes unavailable:

\begin{equation}
    x_j^{\text{dropped}}(t) = 0 \quad \Rightarrow \quad \Delta \phi_{i,j}\text{ is unavailable.}
\end{equation}

This results in a loss of spatial diversity, which negatively impacts AoA estimation accuracy. The expected error in AoA estimation increases with the number of dropped antennas. Since the model can no longer learn a complete phase difference structure across antennas, removing antennas significantly degrades AoA estimation.

\subsubsection{Task-Aware Training Algorithm}
To align signal transformations with feature-specific learning objectives, we design a contrastive training algorithm that selects appropriate augmentations based on the targeted features: temporal patterns (as in modulation classification), spatial phase relationships (as in AoA prediction), or a combination for joint feature learning. Each input sample is augmented twice using the relevant augmentation set, and the resulting views are passed through a shared encoder and projection head, as illustrated in Figure~\ref{fig:frameworkmethod}. The contrastive loss is computed across all positive and negative pairs in the embedding space, encouraging similarity between views of the same sample and dissimilarity across distinct samples.

Algorithm~\ref{alg:task_simclr} outlines the complete training pipeline, emphasizing how the augmentation policy depends on the targeted features, consistent with the self-supervised pretraining and fine-tuning phases depicted in Figure~\ref{fig:framework} and ~\ref{fig:frameworkmethod}.

In addition to these feature-aware augmentations, we apply general-purpose transformations such as amplitude scaling and additive Gaussian noise for regularization. Specifically, amplitude scaling perturbs the signal using a random factor drawn uniformly from the range\([-0.1, 0.1]\), while additive Gaussian noise is applied with a standard deviation of 0.09. While these transformations are not explicitly designed to preserve spatial or temporal structure, they improve generalization and robustness without disrupting the feature-aware representation learning process.

\SetKwFor{ForEach}{for each}{do}{end}

\begin{algorithm}
\caption{Task-Aware SimCLR Training Algorithm}
\SetKwInOut{Input}{Input}
\SetKwInOut{Output}{Output}
\label{alg:task_simclr}
\Input{Batch size $N$, temp. $\tau$, encoder $f$, proj. head $g$, task ID $\text{TASK} \in \{1,2,3\}$, aug. probs $(\alpha, \beta, \gamma)$}
\Output{Trained encoder $f(\cdot)$}

\Repeat{convergence is reached}{
  \ForEach{minibatch $\{x_k\}_{k=1}^{N}$}{
    \ForEach{$k \in \{1, \dots, N\}$}{
        \tcp*{Select Task-Spec. Augm.}
        \uIf{$\text{TASK} = 1$ (AoA)}{
            $t, t' \sim \mathcal{T}_{\text{AoA}} = \{\text{Mask.}(\alpha), \text{T.Shift}(\gamma)\}$\;
        }
        \ElseIf{$\text{TASK} = 2$ (Mod)}{
            $t, t' \sim \mathcal{T}_{\text{Mod}} = \{\text{Drop.}(\beta), \text{T.Shift}(\gamma)\}$\;
        }
        \ElseIf{$\text{TASK} = 3$ (Joint)}{
            $t, t' \sim \mathcal{T}_{\text{Joint}} = \{\text{Mask.}(\alpha), \text{Drop.}(\beta), \text{T.Shift}(\gamma)\}$\;
        }
        \tcp*{Apply Augm. and Encode}
        $\tilde{x}_{2k-1} \gets t(x_k)$; \quad $h_{2k-1} \gets f(\tilde{x}_{2k-1})$; \quad $z_{2k-1} \gets g(h_{2k-1})$\;
        
        $\tilde{x}_{2k} \gets t'(x_k)$; \quad $h_{2k} \gets f(\tilde{x}_{2k})$; \quad $z_{2k} \gets g(h_{2k})$\;
    }
    \tcp*{Compute Cosine Similar.}
    \For{$i,j \in \{1, \dots, 2N\}$}{
        $s_{i,j} \gets \frac{z_i \cdot z_j}{\|z_i\| \|z_j\|}$\;
    }
    \tcp*{Compute InfoNCE Loss}
    $\ell(i,j) \gets - \log \frac{\exp(s_{i,j}/\tau)}{\sum_{k \neq i} \exp(s_{i,k}/\tau)}$\;
    $\mathcal{L} \gets \frac{1}{2N} \sum_{k=1}^{N} \left[ \ell(2k{-}1, 2k) + \ell(2k, 2k{-}1) \right]$\;
    
    \tcp*{Update Networks}
    Optimize $f, g$ to minimize $\mathcal{L}$ via backpropagation\;
  }
}
\Return{Trained encoder $f(\cdot)$}
\end{algorithm}

\begin{figure}[t]
    \centering
    \includegraphics[width=0.96\linewidth,height =0.45\linewidth]{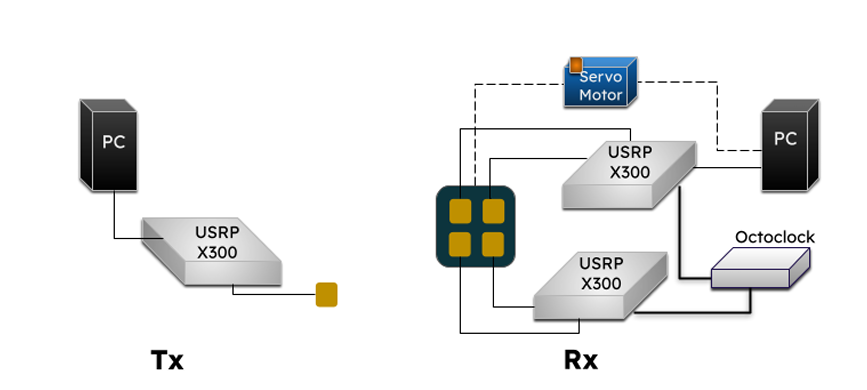}
    \caption{A diagram of the testbed used to create the study dataset.}
    \label{fig:hw}
\end{figure}

\subsection{Datasets}
\label{sec:datasets}
The foundational model was pre-trained on a multi-channel(MIMO), multi-task raw IQ dataset collected over-the-air using a custom testbed. The testbed consisted of a USRP X300 transmitter and two synchronized USRP X300 receivers spaced 2.5 meters apart, each connected to microstrip planar antennas operating at 5.88GHz \ref{fig:hw}. The transmitter used a single antenna, while the receivers employed four antennas in total, divided evenly between the two USRPs. Clock synchronization was achieved using an OctoClock distribution module to ensure coherent multi-channel capture. The dataset includes five digital modulation schemes (16-QAM, 64-QAM, BPSK, QPSK, PAM4) and two continuous waveforms (sine, continuous wave), captured at 1MSps and 10~MSps across three outdoor sessions under realistic multipath conditions. AoA labels were generated by mechanically rotating the transmitter across azimuth and elevation angles from $-70^\circ$ to $70^\circ$ in $10^\circ$ increments, producing 225 classes. Each recording was segmented into fixed-size IQ samples of shape $(4,2,256)$, corresponding to four receive channels, in-phase and quadrature components, and 256 time-domain samples. The dataset contains 1,115,378 labeled samples, with a stratified 70/30 train/test split. This dataset serves both as the pre-training resource for the foundational model and for evaluating its in-distribution, multi-task performance under few-shot adaptation.

To evaluate the model’s ability to generalize beyond the pre-training domain, we conduct experiments on three additional downstream datasets:
\begin{itemize}
\item \textbf{RML2016.10a} for modulation classification. This benchmark dataset contains IQ samples from 11 digital and analog modulation schemes, recorded across SNR levels from –20dB to +18dB. Each class includes 1,000 examples per SNR, with 128 complex-valued IQ samples per example. RML2016.10a introduces unseen modulation types and noise conditions, testing the model’s robustness to domain shifts.
\item \textbf{POWDER RF Fingerprinting dataset} for device identification. This dataset was collected over-the-air from four base stations in the POWDER platform, each transmitting IEEE 802.11a (WiFi) signals using USRP X310 radios. Signals were received by a USRP B210 at 2.685GHz and sampled at 5MS/s. Five 2-second recordings were captured per link per day across two days. This dataset evaluates the model’s ability to distinguish between transmitters in realistic hardware and channel conditions.
\item \textbf{DeepBeam 5-Beam Multi-RF-Chain dataset} for beam prediction. Collected using a multi-RF-chain mmWave testbed at 58 GHz, this dataset includes two transceivers equipped with 4-element patch antenna arrays and individual RF chains, enabling digital beamforming across five predefined beams. For our experiments, we use the dataset as provided, where the IQ data reflects the digitally combined beamformed signal across the four RF chains. We extract single-channel IQ samples of shape $(1,2,256)$ by selecting the first 256 samples from each 1024-sample segment.
\end{itemize}

Together, these datasets offer diverse signal structures, enabling a comprehensive evaluation of the foundational model’s generalization to both in- and out-of-distribution tasks. Table~\ref{tab:datasets} summarizes the datasets used.
\begin{table}[t]
    \centering
    \caption{\textbf{Datasets Used for Pre-Training and Evaluation}}
    \label{tab:datasets}
    \renewcommand{\arraystretch}{1.2}
    \setlength{\tabcolsep}{3.5pt}
    \begin{tabular}{lcc}
        \toprule
        \textbf{Dataset} & \textbf{Sample Shape} & \textbf{Classes} \\
        \midrule
        SSL Pre-Training (Testbed) & (4,2,256) & 7 Mod, 225 AoA \\
        RML2016.10a\cite{rml16} & (1,2,128) & 11 Mod \\
        POWDER RF Fingerprint\cite{powder} & (1,2,256) & 4 Devices \\
        DeepBeam 5-Beam\cite{deepbeam} & (1,2,256) & 5 Beams \\
        \bottomrule
    \end{tabular}
\end{table}

\section{Results and Discussion}

This section presents a comprehensive evaluation of the proposed IQ Wireless Foundation Model across both in-distribution and out-of-distribution tasks. The model is assessed under low-label regimes using frozen encoders and lightweight adaptation strategies, including linear probes and LoRA. Our results highlight how well the model generalizes across tasks and datasets, and how augmentation strategies influence the learning of task-relevant representations.

We begin by describing the encoder architecture, training configuration, and experimental setup. We then evaluate the effectiveness of task-aware augmentations for modulation and AoA tasks, each relying on distinct signal characteristics, temporal and spatial, respectively. In the in-distribution setting, we compare feature-specific training to joint-task training (IQ FM model augmentation strategy) under varying label budgets to analyze performance trade-offs.

Following this, we present results on out-of-distribution datasets and tasks including beam prediction, RF fingerprinting, and modulation classification on RML2016a, showcasing the model’s generalization capability. We then study how the strength of augmentations, such as cyclic time shifting, channel masking, and channel dropping, influences downstream accuracy by biasing the learned representations toward either temporal or spatial characteristics. To further interpret the model’s behavior, we analyze the learned feature space using PCA visualizations and clustering metrics. Finally, we present an ablation study quantifying the contribution of individual augmentations to the model’s overall performance.

\begin{figure}[!b]
    \centering
    \includegraphics[width=0.95\linewidth]{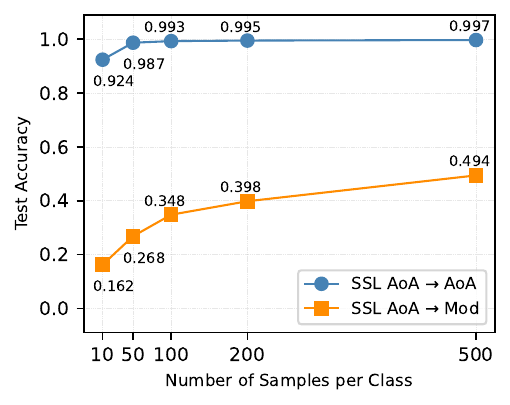}
    \caption{Classification accuracy of an SSL encoder trained with AoA-guided augmentations (TR+CM) when evaluated on AoA and modulation tasks.}
    \label{fig:aoa_aug}
\end{figure}
\begin{table}[t]
    \centering
    \caption{\textbf{Augmentation and Training Parameters for SSL Encoders}}
    \label{tab:ssl_params}
    \renewcommand{\arraystretch}{1.2}
    \setlength{\tabcolsep}{4pt}
    \begin{tabular}{lccc}
        \toprule
        \textbf{Parameter} & \textbf{SSL-Mod} & \textbf{SSL-AoA} & \textbf{SSL-Joint} \\
        \midrule
        CD Probability (\%)           & 100   & 0    & 45  \\
        CM Length                     & 0    & 200 (max)     & 40   \\
        CM Probability (\%)           & 0    & 95  & 97  \\
        TR Length                     & 40   & 120   & 20  \\
        TR Probability (\%)           & 80   & 95   & 95  \\
        Learning Rate                 & 0.1 & 0.1 & 0.5 \\
        Temperature ($\tau$)           & 1.5 & 1.5 & 0.12 \\
        \bottomrule
    \end{tabular}
\end{table}
\subsection{Encoder Architecture and Experimental Setup}

We employ the ShuffleNetV2 (0.5×) architecture as the encoder backbone due to its lightweight structure and suitability for deployment on edge devices. Our implementation, based on the official PyTorch release, contains approximately 341k trainable parameters and leverages depthwise separable and grouped convolutions to optimize for inference efficiency. The encoder is followed by a task-specific projection head during contrastive pretraining, and lightweight classification heads are attached during the fine-tuning phase.

All training and evaluation experiments were conducted on a workstation equipped with a 12th Gen Intel Core i9-12900K CPU, 64~GB of RAM, and an NVIDIA RTX 3080 Ti GPU.
Prior to training, all IQ data samples were normalized using unit max scaling:

\[
\texttt{iq\_data} = \frac{\texttt{iq\_data}}{\max\left( \left| \texttt{iq\_data} \right| \right)}
\]

All models are trained using the AdamW optimizer with a cosine annealing learning rate schedule that decays the learning rate to $1 \times 10^{-7}$ over the training epochs. The augmentation and training parameters for each encoder type are summarized in Table~\ref{tab:ssl_params}.


\begin{figure}[!b]
    \centering
    \includegraphics[width=0.95\linewidth]{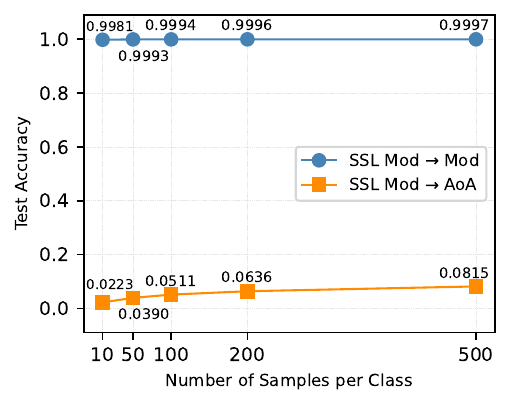}
    \caption{Classification accuracy of an SSL encoder trained with modulation-guided augmentations (TR+CD) when evaluated on modulation and AoA tasks.} 
    \label{fig:mod_aug}
\end{figure}


\subsection{Effectiveness of Augmentation Strategies}

We evaluate the impact of task-specific augmentations on downstream performance using linear probing, where a shallow classifier is trained on top of frozen SSL encoders. Classification accuracy is used as the primary metric, as it reflects both task performance and the quality of learned feature representations.

Figure~\ref{fig:mod_aug} shows results when the encoder is trained with modulation-oriented augmentations, cyclic time shifting (TR) and channel dropping (CD). This combination emphasizes temporal structure in the signal, enabling the encoder to capture features relevant to symbol timing and transitions. As a result, the model achieves 99.81\% modulation accuracy with only 10 labeled samples per class, and up to 99.97\% with 500 samples. However, this encoder performs poorly on AoA classification, reaching just 8.15\% accuracy, which highlights a mismatch between temporal augmentations and the spatial dependencies required for angle estimation.

In contrast, Figure~\ref{fig:aoa_aug} reports results for encoders trained with AoA-oriented augmentations, TR combined with channel masking (CM), which preserves inter, antenna spatial structure. This strategy yields 92.4\% accuracy on AoA classification with only 10 samples per class, across 225 angle classes. However, the same encoder performs poorly on modulation classification (16.2\% accuracy), showing a 6.16× drop compared to the modulation-specific encoder. These results illustrate how different augmentations bias the encoder toward learning task-relevant but non-transferable features.

Overall, these findings confirm that pairing augmentations with the dominant signal characteristics of a given task, temporal for modulation and spatial for AoA, is critical for effective SSL. Channel dropping encourages temporal sensitivity, whereas channel masking preserves spatial coherence across antenna elements. While task-specific augmentations guide the encoder toward specialized features, they limit cross-task generalization. This limitation motivates the need for a unified augmentation strategy that supports joint feature learning, which we explore in the next section.


\begin{figure}[!t]
    \centering
    \includegraphics[width=0.95\linewidth]{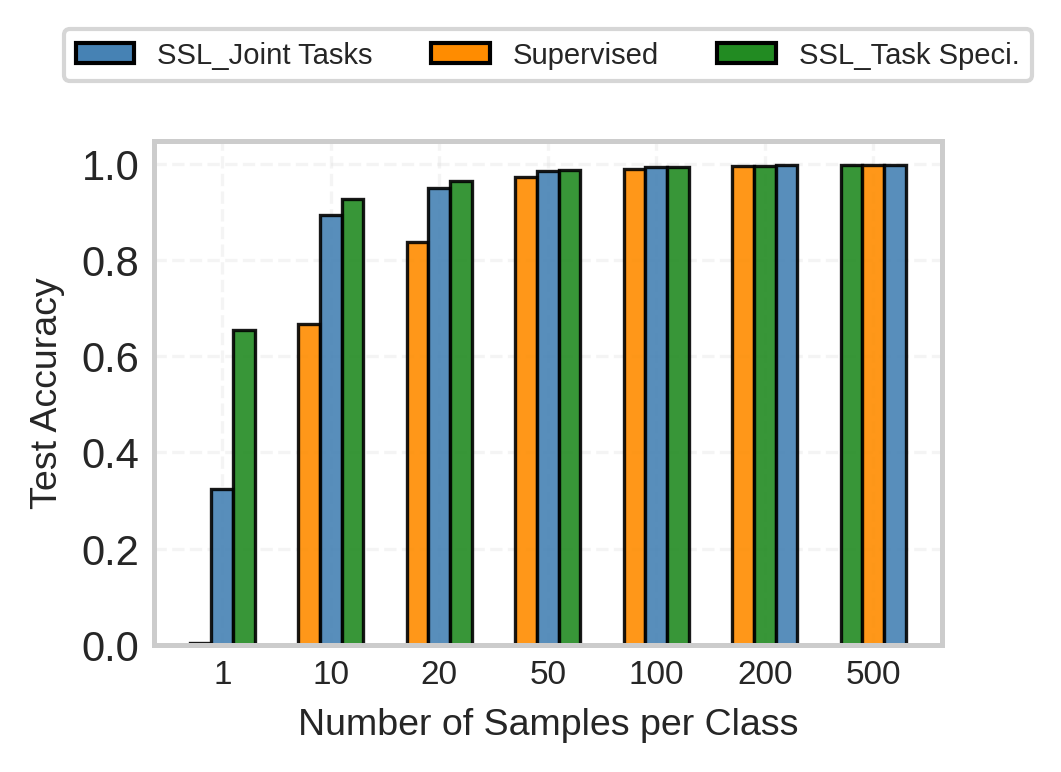}
    \caption{AoA classification accuracy for supervised learning, task-specific SSL, and joint-task SSL across varying numbers of labeled samples per class.}
    \label{fig:aoa_joint}
\end{figure}

\begin{figure}[!t]
    \centering
    \includegraphics[width=0.95\linewidth]{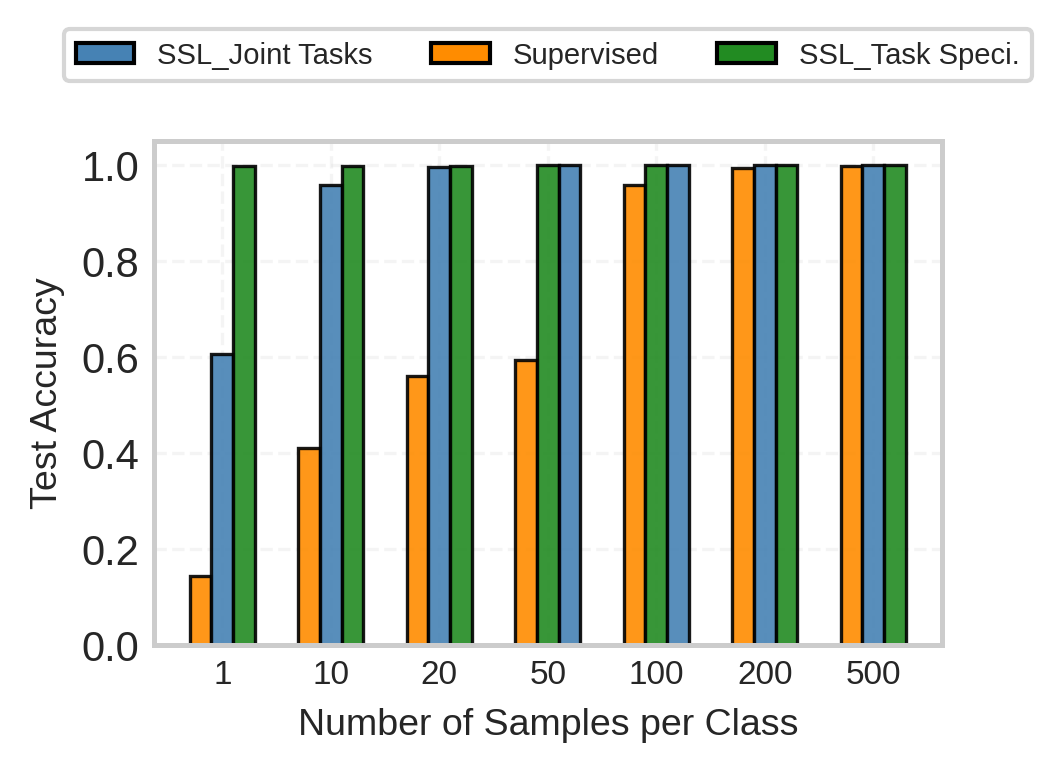}
    \caption{Modulation classification accuracy for supervised learning, task-specific SSL, and joint-task SSL across varying numbers of labeled samples per class.}
    \label{fig:mod_joint}
\end{figure}

\subsection{In-Distribution Performance: I/Q FM vs. Feature-Specific}

We evaluate the IQ Wireless Foundation Model on in-distribution tasks, where both modulation and AoA classification are drawn from the same dataset used for SSL pretraining. Here, the model is trained with joint task-aware augmentations, without access to task labels, allowing it to learn both temporal and spatial features within a unified representation space.

We compare three approaches: (i) the I/Q FM encoder trained with joint-task augmentations (referred to as “Joint-task” in figures), (ii) task-specific SSL encoders optimized separately for modulation or AoA, and (iii) a fully supervised baseline. All models are evaluated under varying label budgets using frozen encoders and linear probes.

Figures~\ref{fig:aoa_joint} and~\ref{fig:mod_joint} show that task-specific SSL consistently outperforms both joint-task training and the supervised baseline in low-shot scenarios. For example, in the one-shot setting, task-specific SSL achieves 65.45\% accuracy for AoA—145$\times$ better than the supervised model (0.45\%) and 2$\times$ higher than the joint model (32.42\%). For modulation classification, it reaches 99.67\%, surpassing the supervised baseline (14.27\%) and the joint model (60.48\%) by 6.98$\times$ and 1.36$\times$, respectively.

Table~\ref{tab:ssl_accuracy} summarizes accuracy across label budgets. Task-specific SSL retains an advantage up to 100 samples per class, after which performance converges across all models. For fairness, the same label subsets and learning rates were used: \(10^{-2}\) for supervised and \(10^{-3}\) for SSL fine-tuning.

While task-specific encoders offer superior performance in low-label regimes, they lack the flexibility of a unified model. The IQ FM, trained via joint augmentations, learns both spatial and temporal features, making it more generalizable and better suited for multi-task scenarios, as explored in the subsequent section.

\begin{table}[b]
    \centering
    \scriptsize 
    \renewcommand{\arraystretch}{1.1} 
    \setlength{\tabcolsep}{2pt} 
    \caption{\textbf{Test Accuracy for Different Sample Sizes in Two SSL Tasks (\%)}}
    \label{tab:ssl_accuracy}

    \resizebox{\columnwidth}{!}{ 
    \begin{tabular}{c c c c c}
        \toprule
        \textbf{\makecell{\# labeled \\ Samples/Class}} & \textbf{Task} & \textbf{\makecell{SSL\\Joint Tasks}} & \textbf{Supervised} & \textbf{\makecell{SSL \\Task Specific}} \\
        \midrule
        \multirow{2}{*}{1}   & Mod & 60.48   & 14.27   & \textbf{99.67}   \\ 
                             & AoA & 32.42   & 0.45    & \textbf{65.45}   \\ 
        \midrule
        \multirow{2}{*}{10}  & Mod & 95.71   & 41.07   & \textbf{99.81}   \\ 
                             & AoA & 89.35   & 66.58   & \textbf{92.45}    \\ 
        \midrule
        \multirow{2}{*}{100} & Mod & \textbf{99.95} & 95.66   & 99.94   \\ 
                             & AoA & 99.33 & 98.81   & \textbf{99.34}   \\ 
        \midrule
        \multirow{2}{*}{500} & Mod &  \textbf{99.97}  & 99.75   & \textbf{99.97}   \\ 
                             & AoA & \textbf{99.76}   & 99.73   & 99.72   \\ 
        \bottomrule
    \end{tabular}
    }
\end{table}


\begin{figure*}[t]
    \centering
    \subfloat[RF fingerprinting\label{fig:rf_fingerprint}]{
        \includegraphics[height=5.1cm, width=0.31\textwidth]{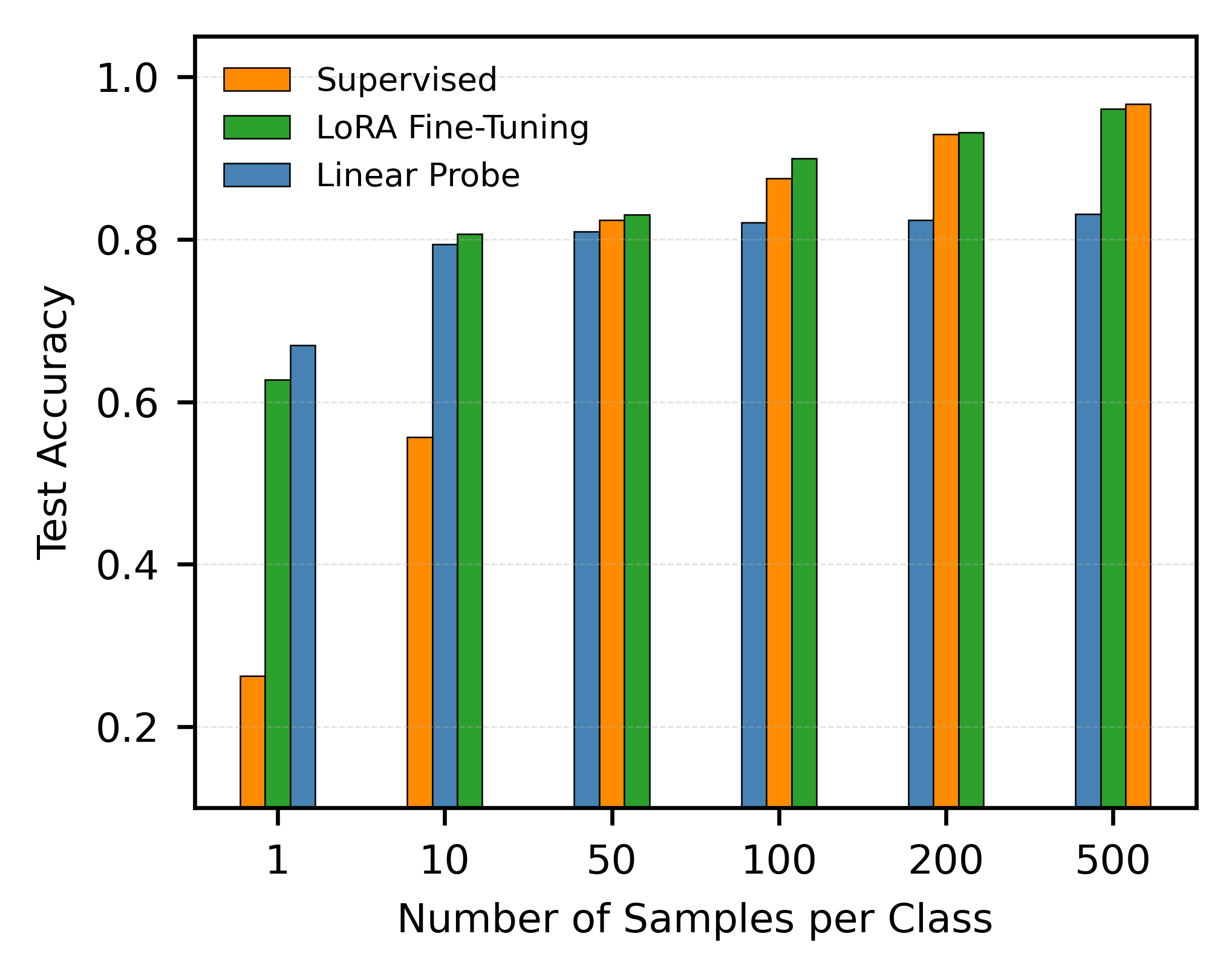}
    }
    \hfill
    \subfloat[RML16\label{fig:rml16}]{
        \includegraphics[height=5.1cm,width=0.31\textwidth]{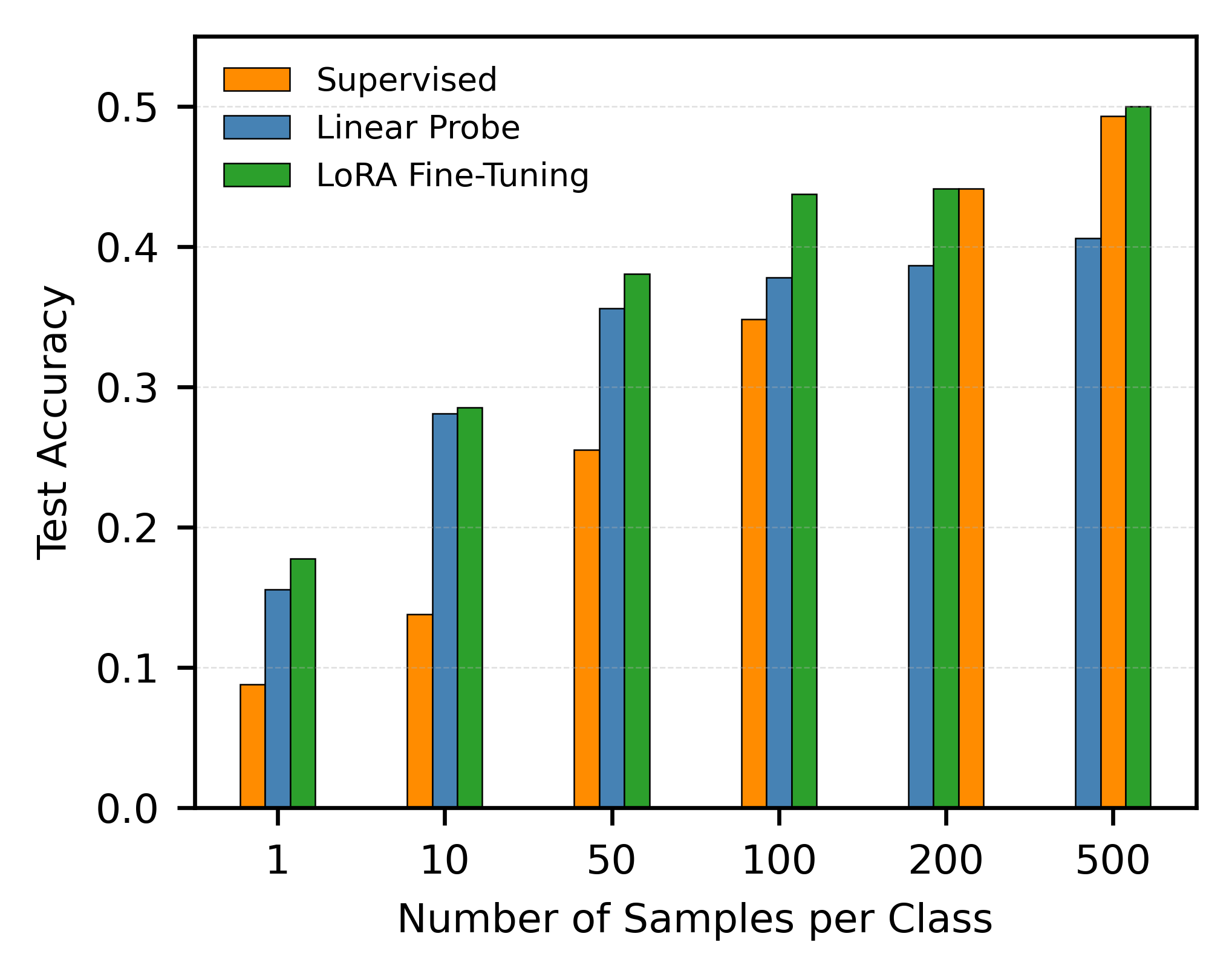}
    }
    \hfill
    \subfloat[DeepBeam\label{fig:deepbeam}]{
        \includegraphics[height=5.1cm, width=0.31\textwidth]{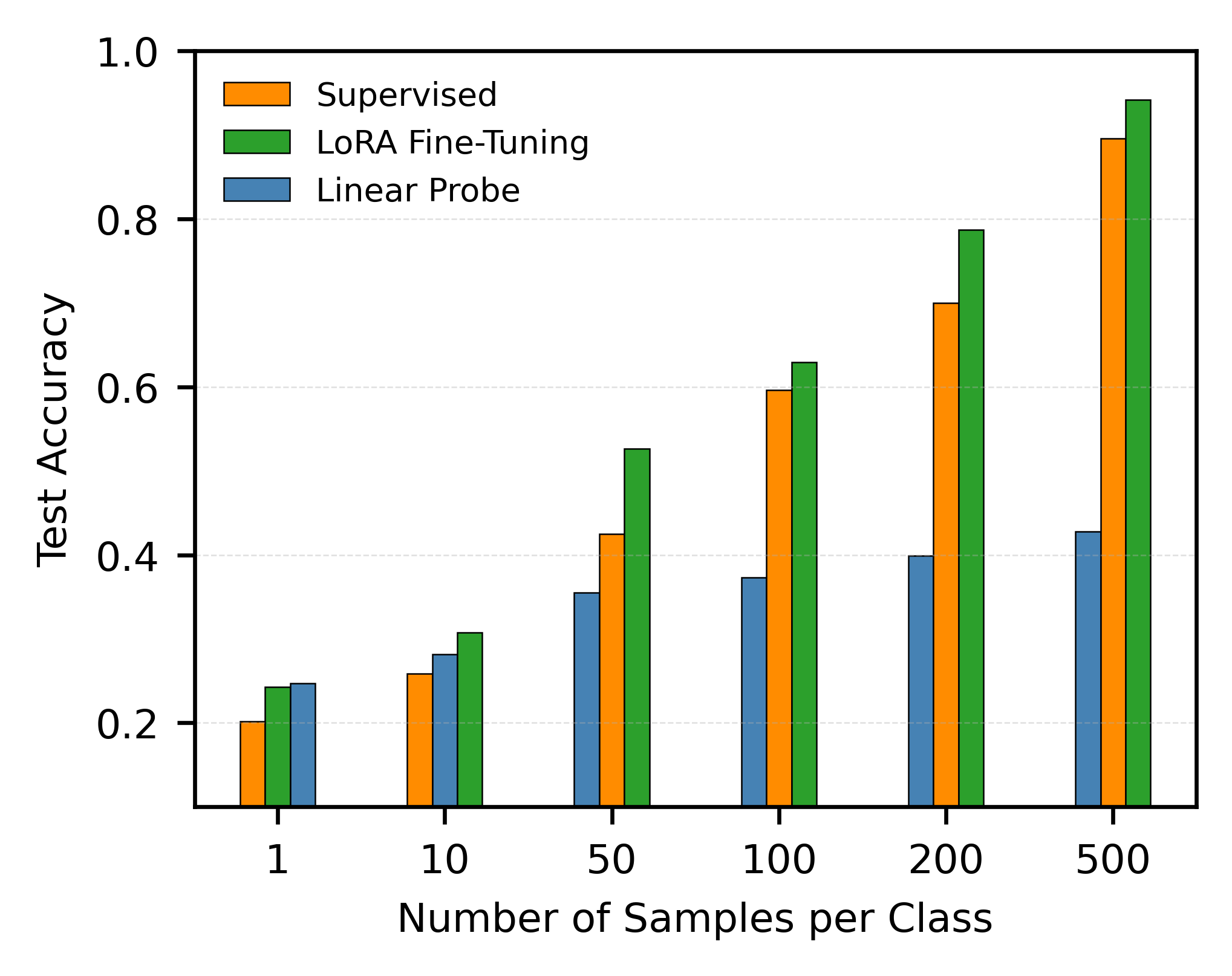}
    }
    
    \caption{Test accuracy of different models across varying number of samples per class on (a) RF fingerprinting, (b) RML16, and (c) DeepBeam datasets.}
    \label{fig:all_datasets}
\end{figure*}

\subsection{Evaluation Results on Multiple Datasets}

We now evaluate the generalization performance of the IQ Wireless Foundation Model (IQ FM) on three out-of-distribution datasets introduced in Section~\ref{sec:datasets}. RF fingerprinting and beam prediction represent entirely new tasks not seen during SSL pretraining, while \textsc{RML2016} serves as a modulation classification benchmark with unseen modulation types and significantly lower SNR conditions. These datasets vary in signal structure, channel count, sample length, and label space, allowing us to isolate the model’s robustness to both task and domain shift.

To ensure input compatibility, RF fingerprinting and DeepBeam signals are segmented into 256-sample frames with missing channels padded by zeros, while the single-channel 128-sample \textsc{RML2016} samples are processed via the encoder’s global pooling layer. For all experiments, the encoder remains frozen, and two lightweight adaptation strategies are evaluated: (i) linear probing, which appends a linear classifier on top of the frozen encoder, and (ii) Low-Rank Adaptation, which introduces approximately 84K trainable parameters into the convolutional layers. In contrast, the fully supervised baseline trains the entire ShuffleNetV2 (0.5×) backbone comprising 341.8K parameters, excluding the final task-specific linear layer whose size varies with the number of output classes.

All models are trained for 100 epochs. For linear probing, the classifier uses a learning rate of $1 \times 10^{-3}$, while LoRA and supervised models use $1 \times 10^{-2}$. LoRA is configured with rank $r=1$ and task-specific scaling factors: $\alpha=35$ for RF fingerprinting, $\alpha=10$ for DeepBeam, and $\alpha=60$ for RML16 (reduced to $\alpha=50$ at 200 and 500 samples per class). Batch sizes are selected empirically for each dataset to optimize performance.

The results, summarized in Figure~\ref{fig:all_datasets}, demonstrate that the IQ FM encoder generalizes well across unseen tasks and domains. For RF fingerprinting (Figure~\ref{fig:rf_fingerprint}), LoRA achieves 62.8\% accuracy with a single labeled sample per class, outperforming the supervised baseline (26.3\%) and closely matching linear probing (66.9\%). At 500 samples per class, LoRA reaches 96.0\%—nearly matching supervised accuracy (96.6\%) while using an order of magnitude fewer parameters.

On the \textsc{RML2016} benchmark (Figure~\ref{fig:rml16}), LoRA shows strong performance despite the domain shift. It achieves 17.8\% accuracy with one labeled sample per class, compared to 15.6\% for linear probing and only 8.8\% for supervised training. At 500 samples per class, LoRA reaches 50.0\% accuracy, outperforming both the supervised baseline (49.3\%) and linear probing (40.6\%).

For beam prediction (Figure~\ref{fig:deepbeam}), LoRA again provides a consistent advantage. With 50 labeled samples per class, it achieves 52.6\% accuracy, compared to 42.5\% for the supervised model and 35.5\% for linear probing. At 500 samples per class, LoRA achieves 94.1\%, closely matching supervised training (89.5\%), while linear probing trails at 42.7\%.

Overall, these results highlight the strong out-of-distribution generalization capability of the IQ FM model. LoRA offers a lightweight and effective adaptation mechanism, achieving near-supervised performance across diverse tasks while keeping the encoder frozen and minimizing parameter overhead.




\subsection{Impact of Augmentation Probabilities on Performance}

\begin{figure}[t]
    \centering
    \includegraphics[width=0.98\linewidth]{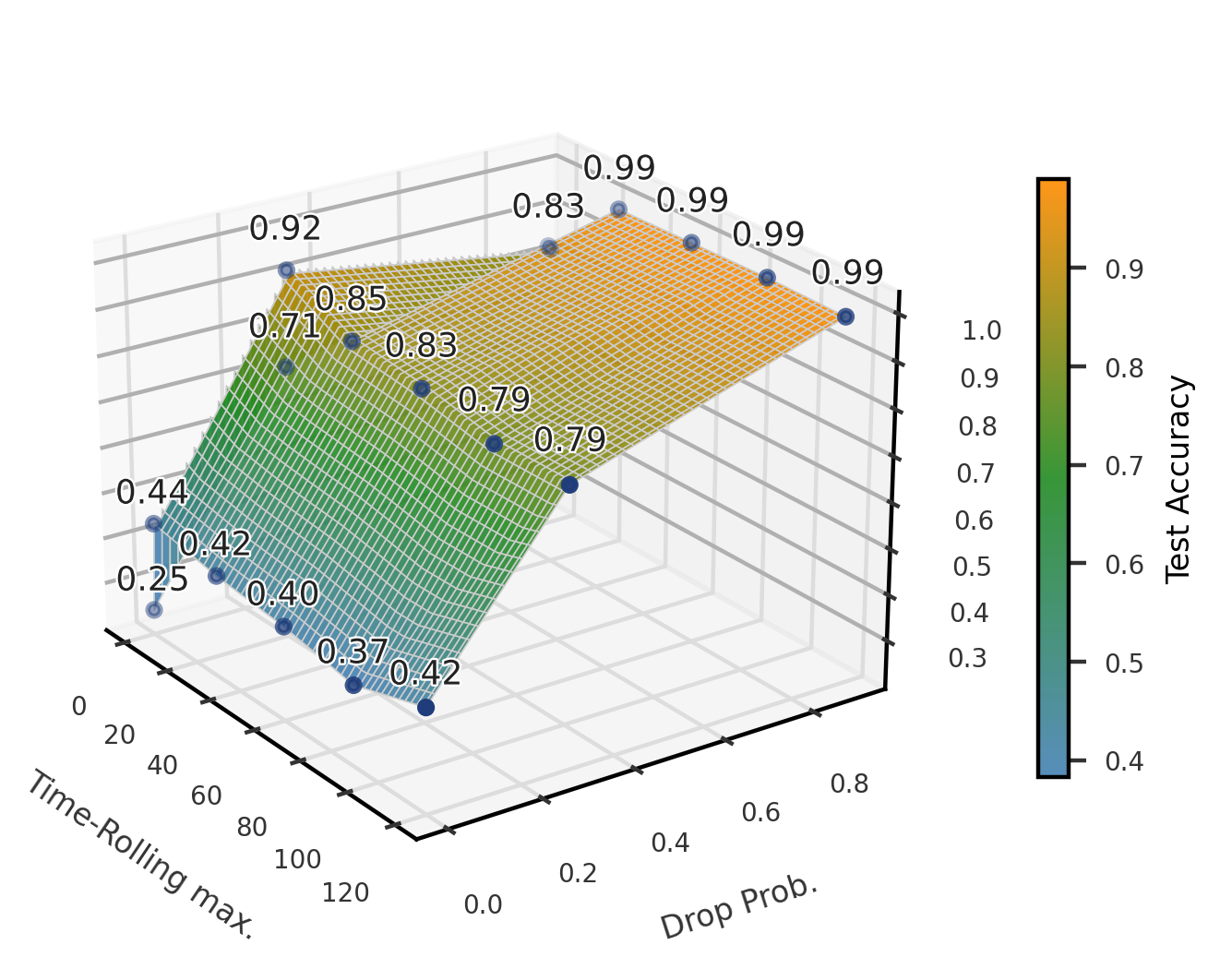}
    \caption{Surface plot illustrating the effect of (TR \& CD) on modulation classification accuracy.}
    \label{fig:surface_plot_mod}
\end{figure}

\begin{figure}[t]
    \centering
    \includegraphics[width=0.98\linewidth]{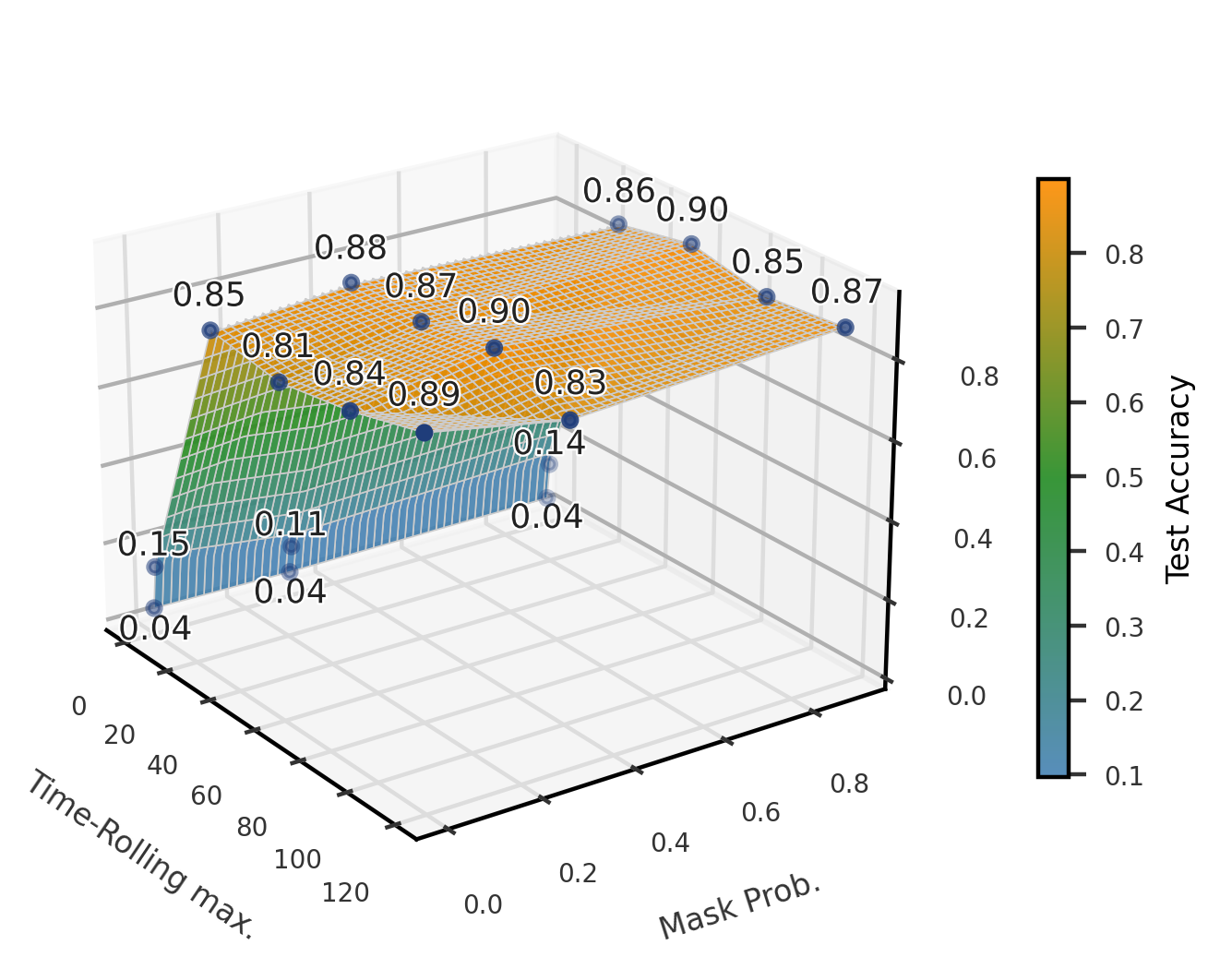}
    \caption{Surface plot depicting the influence of (TR \& CM) on AoA classification accuracy}
    \label{fig:surface_plot_aoa}
\end{figure}

To investigate how augmentation parameters influence self-supervised representation quality, we systematically varied the probabilities of channel dropping (or masking) and the magnitude of time rolling. The resulting effects on downstream task performance were visualized using smoothed surface plots generated via interpolation, as shown in Figures~\ref{fig:surface_plot_mod} and~\ref{fig:surface_plot_aoa}. Each surface plot corresponds to an encoder trained with contrastive SSL for 20 epochs, followed by linear probing using 50 labeled samples per class.

Figure~\ref{fig:surface_plot_mod} illustrates that modulation classification is highly sensitive to the probability of channel dropping. For instance, when the maximum time rolling is fixed at 60 samples, increasing the dropping probability from 1\% to 90\% improves classification accuracy from 40\% to 99\%. The plot also reveals that time rolling is essential as the modulation accuracy remains below 85\% when time rolling is limited to fewer than 30 samples, even at high dropping rates.

In contrast, Figure~\ref{fig:surface_plot_aoa} shows that AoA classification is less sensitive to channel masking. Nonetheless, performance still improves moderately, with the time rolling parameter fixed at 60 samples, increasing the masking probability from 1\% to 90\% raises accuracy from 81\% to 90\%. This suggests that while channel masking supports spatial feature learning, its impact is less pronounced than channel dropping in the modulation task.

These findings confirm that time rolling serves as a core augmentation for both tasks, while channel dropping and channel masking act as task-specific refinements. The surface plots further demonstrate that augmentation probability, and thus transformation intensity, directly influences the effectiveness of self-supervised learning. To further validate the role of time rolling, we present an ablation study in the following subsection, quantifying its individual and combined effects with task-specific augmentations.

\subsection{Clustering Behavior via PCA and Pseudo-Labeling}

\begin{figure*}[tb]
    \centering
    \begin{subfigure}{0.31\textwidth}
        \centering
        \includegraphics[width=\linewidth]{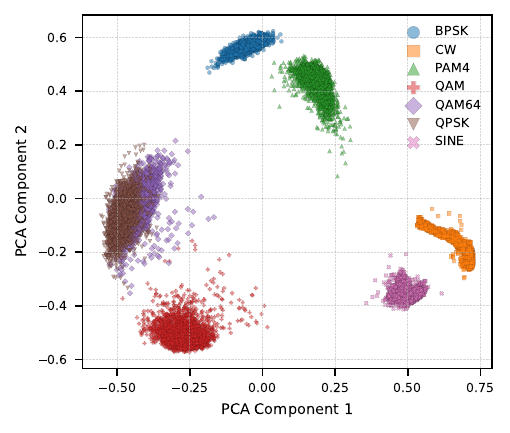}
        \caption{}
        \label{fig:mod_clusters}
    \end{subfigure}
    \hfill
    \begin{subfigure}{0.31\textwidth}
        \centering
        \includegraphics[width=\linewidth]{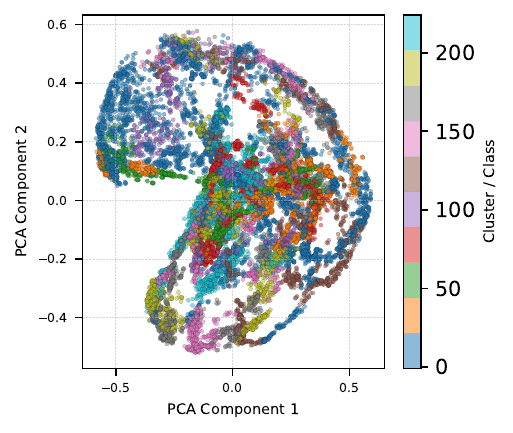}
        \caption{}
        \label{fig:aoa_clusters}
    \end{subfigure}
    \hfill
    \begin{subfigure}{0.31\textwidth}
        \centering
        \includegraphics[width=\linewidth]{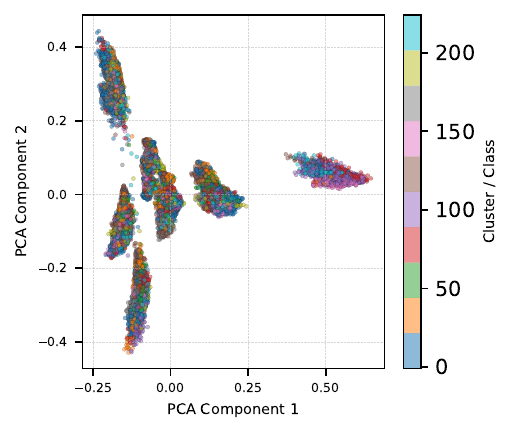}
        \caption{}
        \label{fig:aoa_mod_clusters}
    \end{subfigure}
    \caption{PCA visualization of learned feature representations for (a) modulation-specific, (b) AoA-specific, and (c) joint-task SSL models. Points are colored by pseudo-label clusters, with class legends (a) and color bars (b, c) reflecting the number of classes.}
    \label{fig:combined_clusters}
\end{figure*}

To further assess the impact of augmentation strategies on representation learning, we examine the clustering behavior of three trained SSL models, modulation-specific, AoA-specific, and joint-task, using PCA projections and KNN-based pseudo-labeling. Figure~\ref{fig:combined_clusters} presents the PCA visualizations, highlighting distinct patterns in feature separability across the models.

Pseudo-labeling is performed via KNN clustering on features extracted from 20k unlabeled samples. Cluster centroids are computed, and labels are propagated from a single labeled sample per class to all samples in the corresponding cluster based on proximity. This procedure enables a fully self-supervised evaluation, without fine-tuning or weight updates, of the encoder’s ability to produce semantically meaningful representations.

The modulation-specific SSL encoder (Figure~\ref{fig:mod_clusters}) forms seven well-separated clusters, corresponding precisely to the seven modulation classes. This demonstrates that modulation-guided augmentations enhance temporal feature extraction and facilitate class-level separability. As a result, pseudo-labeling achieves 99.99\% accuracy (Table~\ref{tab:pseudo_labeling}), confirming the encoder’s effectiveness in modulation-specific representation learning.

In contrast, the AoA-specific SSL encoder (Figure~\ref{fig:aoa_clusters}) organizes the feature space primarily by spatial characteristics. Unlike the joint-task model, which exhibits clear grouping into seven modulation-driven clusters, the AoA-specific encoder produces more dispersed and irregular clusters that do not reflect any modulation-based structure. This distribution suggests that the learned features are shaped by angle-of-arrival variations rather than temporal modulation patterns. Due to the fine angular resolution (10$^\circ$ steps), clusters exhibit higher intra-class variance, limiting the pseudo-labeling accuracy to 57.36\%. Further analysis using silhouette scores is presented in the next subsection, as the precise number of clusters formed by KNN is difficult to infer directly from the PCA projection. When evaluated on modulation classification, this encoder achieves only 10.62\%, reinforcing its task-specific nature and limited generalization across tasks.

The joint-task encoder (Figure~\ref{fig:aoa_mod_clusters}) exhibits a hierarchical clustering structure, with seven dominant global clusters clearly visible in the PCA plot. These clusters correspond to different modulation types, reflecting how PCA captures global relationships in the latent space. Notably, the encoder achieves 99.93\% and 43.61\% pseudo-labeling accuracy for modulation and AoA classification, respectively. This suggests the presence of localized clustering within each modulation group based on angle-of-arrival. Joint-task training thus enables the model to encode both temporal and spatial features, global separation by modulation and fine-grained organization by AoA within each cluster. While this entangled structure makes pseudo-labeling more challenging than with task-specific models, it also reveals the model's capacity to learn multi-level semantic structure from unlabeled data.

\subsection{Cluster Quality Assessment via Silhouette Score}
\begin{table}[b]
    \centering
    \caption{\textbf{Silhouette-Based Clustering and Pseudo-Labeling Accuracy for Task-Specific and Joint SSL Encoders}}
    \label{tab:pseudo_labeling}
    \renewcommand{\arraystretch}{1.2}
    \setlength{\tabcolsep}{4pt}
    \begin{tabular}{l c cc}
        \toprule
        \textbf{Model} & \textbf{Optimum Clusters} & \multicolumn{2}{c}{\shortstack{\textbf{Pseudo-Labeling} \\ \textbf{Accuracy (\%)}}} \\
        \cmidrule(lr){3-4}
        & (Silhouette) & \textbf{Modulation} & \textbf{AoA} \\
        \midrule
        SSL AoA Task        & 244  & 10.62         & \textbf{57.36} \\
        SSL Modulation Task & 7    & \textbf{99.99} & 0.53 \\
        SSL Joint-Tasks     & 1600 & \textbf{99.99} & 43.61 \\
        \bottomrule
    \end{tabular}
\end{table}

To quantify cluster compactness and separation, we compute the silhouette coefficient, defined for each sample $i$ as:
\[
s(i) = \frac{b(i) - a(i)}{\max\{a(i), b(i)\}},
\]
where $a(i)$ is the average intra-cluster distance and $b(i)$ is the average distance to the nearest neighboring cluster. The overall silhouette score is given by:
\[
S = \frac{1}{N} \sum_{i=1}^{N} s(i).
\]
Higher values indicate tighter intra-cluster cohesion and stronger inter-cluster separation.

\begin{figure*}[tb]
    \centering
    \begin{subfigure}{0.31\textwidth}
        \centering
        \includegraphics[width=\linewidth]{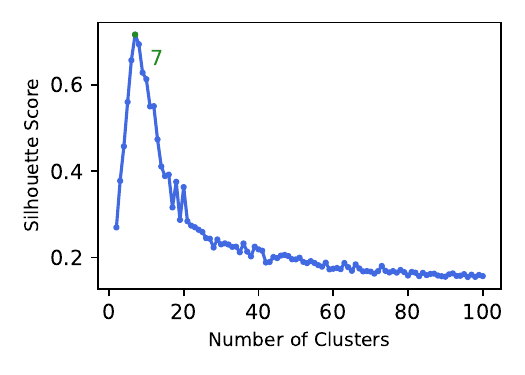}
        \caption{}
        \label{fig:sil_mod_clusters}
    \end{subfigure}
    \hfill
    \begin{subfigure}{0.31\textwidth}
        \centering
        \includegraphics[width=\linewidth]{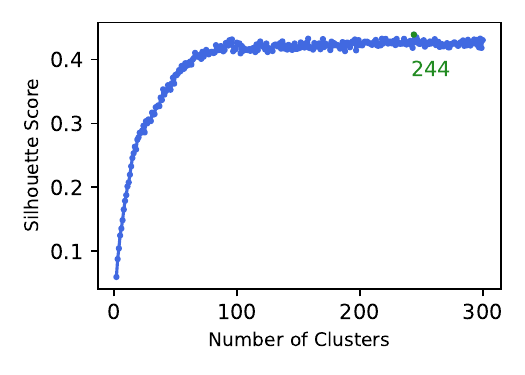}
        \caption{}
        \label{fig:sil_aoa_clusters}
    \end{subfigure}
    \hfill
    \begin{subfigure}{0.31\textwidth}
        \centering
        \includegraphics[width=\linewidth]{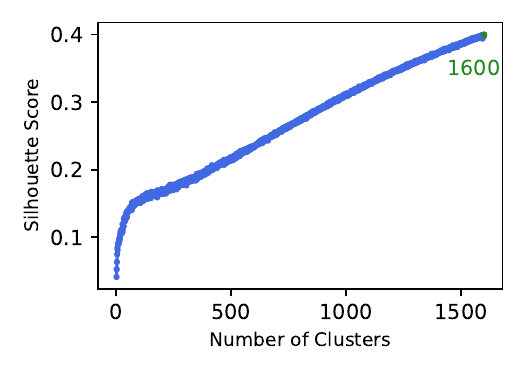}
        \caption{}
        \label{fig:sil_aoa_mod_clusters}
    \end{subfigure}
    \caption{Silhouette score versus number of clusters for three SSL encoders: (a) modulation-specific, (b) AoA-specific, and (c) joint-task. Each plot illustrates the clustering quality of the learned feature space, with peak locations reflecting task-dependent structure.}
    \label{fig:combined_Silhouette}
\end{figure*}

Figure~\ref{fig:combined_Silhouette} shows silhouette scores across the three models. For modulation classification (left), the score peaks when the number of clusters matches the number of modulation classes (7), confirming that the encoder forms well-separated clusters. For AoA (middle), the silhouette score peaks near 244 clusters, which closely aligns with the 225 discrete angle classes. The slight offset is expected due to intra-class variance and overlapping spatial features at small angular resolutions (10$^\circ$), which may cause some neighboring angles to merge or split during unsupervised clustering. In the joint-task case (right), the silhouette score gradually increases and peaks near 1,600 clusters. This aligns with the expected number of distinct semantic combinations:
\[
C_{\text{joint}} = C_{\text{mod}} \times C_{\text{aoa}} = 7 \times 225 = 1575.
\]
The close agreement between the peak cluster count and the theoretical class combinations confirms that the encoder captures both temporal and spatial structure. While PCA primarily reveals the modulation-driven global grouping, the silhouette score reflects the finer AoA sub-structure embedded within each modulation cluster. The cluster counts and corresponding pseudo-labeling accuracies are summarized in Table~\ref{tab:pseudo_labeling}.

These observations are consistent with the trends in Table~\ref{tab:ssl_accuracy}, where the AoA-specific model outperforms the joint-task encoder in AoA classification under low-label regimes. Since clustering occurs independently of modulation, the AoA-specific model generalizes well across modulation types using fewer labeled samples, enabling more effective fine-tuning. In contrast, the joint-task model exhibits a hierarchical structure in which modulation groups form first, followed by sub-clustering based on AoA. This entangled structure requires more labeled data to fine-tune effectively, as decision boundaries must resolve both spatial and temporal dependencies.

However, the spatial structure learned by the AoA-specific model introduces a trade-off. While it enables modulation-agnostic AoA generalization, the resulting clusters are tightly packed, limiting their separability in high-data regimes. As more labeled data becomes available, the joint-task model begins to outperform the AoA-specific model, which struggles due to reduced cluster separability in its latent space. This explains why the AoA-specific encoder excels in low-label scenarios but is ultimately surpassed by the joint-task encoder as supervision increases.


\subsection{Ablation Study on Augmentations}

To quantify the contribution of each augmentation, we conduct an ablation study by selectively applying cyclic time rolling (TR), channel masking (CM), and channel dropping (CD). Table~\ref{tab:augmentation_ablation} reports the classification accuracy under various label budgets. In all cases, the SSL encoder was pretrained for only 10 epochs.

For AoA classification, the combination of TR and CM yields the highest performance, reaching 70.21\% accuracy with 10 labeled samples per class and 99.22\% with 200 samples. Removing CM reduces accuracy to 59.33\%, while excluding both TR and CM results in a substantial decline to 0.95\%, underscoring the critical role of spatial augmentations for angular representation learning.

For modulation classification, the optimal configuration involves TR combined with CD, achieving 93.02\% accuracy at 10 samples per class and 96.98\% at 200 samples. Omitting CD leads to a reduction in performance to 68.7\%, while removing both augmentations further decreases accuracy to only 21.74\%. These results confirm the importance of temporal augmentations in facilitating effective modulation discrimination.

\begin{table}[b]
    \centering
    \scriptsize 
    \renewcommand{\arraystretch}{1.0} 
    \setlength{\tabcolsep}{2pt} 
    \caption{\textbf{Ablation Study of Augmentation Strategies} \\ With Different Sample Sizes (\%)}
    \label{tab:augmentation_ablation}
    
    \resizebox{\columnwidth}{!}{ 
    \begin{tabular}{ccc ccc ccc}
        \toprule
        \multicolumn{3}{c}{\textbf{Augmentations}} & \multicolumn{3}{c}{\textbf{AoA Accuracy (\%)}} & \multicolumn{3}{c}{\textbf{Mod Accuracy (\%)}} \\
        \cmidrule(lr){1-3} \cmidrule(lr){4-6} \cmidrule(lr){7-9}
        \textbf{Time Rolling} & \textbf{Masking} & \textbf{Dropping} & \textbf{10} & \textbf{100} & \textbf{200} & \textbf{10} & \textbf{100} & \textbf{200} \\
        \midrule
        \cmark &  &  & 37.21 & 59.21 & 63.33 & 36.86 & 55.85 & 60.12 \\
        & \cmark &  & 1.22 & 1.89 & 2.16 & 16.85 & 24.11 & 26.28 \\
        &  & \cmark & 0.73 & 0.98 & 1.09 & 21.64 & 24.71 & 25.23 \\
        \cmark & \cmark &  & \textbf{73.58} & \textbf{90.61} & \textbf{92.41} & 24.34 & 37.95 & 44.24 \\
        \cmark &  & \cmark & 5.23 & 12.32 & 15.05 & \textbf{99.67} & \textbf{99.8} & \textbf{99.84} \\
        \cmark & \cmark & \cmark & 36.19 & 65.69 & 70.77 & 87.25 & 95.55 & 96.38 \\
        \bottomrule
    \end{tabular}
    }
\end{table}

\section{Conclusion}

This paper presented the first foundational model for wireless communications that operates directly on raw IQ data, enabling multi-task learning across modulation classification, angle-of-arrival estimation, RF fingerprinting, and beam prediction. The proposed architecture combines a lightweight ShuffleNetV2 (0.5×) backbone encoder with task-specific linear probes or parameter-efficient LoRA, enabling scalable and efficient adaptation to diverse tasks without the need for handcrafted features or extensive preprocessing. A principled, task-aware augmentation strategy was introduced to capture temporal and spatial structures in raw MIMO IQ data, supporting robust representation learning across tasks. Comprehensive evaluations on multiple over-the-air datasets demonstrated the model's strong generalization capabilities: task-specific SSL achieved up to 99.67\% and 65.45\% accuracy for modulation and AoA classification, respectively, with only one labeled sample per class, while the joint SSL model reached 95.71\% and 89.35\% accuracy with 10 samples per class. The foundational model also generalized to unseen tasks and out-of-distribution datasets, achieving 66.9\% accuracy on RF fingerprinting (2.5× supervised), 52.6\% on beam prediction (1.2× supervised), and 38.1\% on RML2016 modulation classification (1.5× supervised) at 50 samples per class. These results demonstrate the potential of raw IQ-based foundational models as scalable, efficient solutions for future AI-native 6G networks.


\bibliographystyle{IEEEtran}
\bibliography{name.bib}

\end{document}